\documentclass[conference,compsoc]{IEEEtran}
\usepackage{epsfig,endnotes}
\usepackage{diagbox,amsmath,amsthm,amssymb,booktabs,comment,graphicx,subfigure,multirow,bm,tikz}
\usepackage{tabularx,array,float,colortbl,lstautogobble,color,zi4,listings,bbm}
\usepackage{bbding} 
\usepackage{algorithmic}
\usepackage{setspace}
\usepackage{enumitem}
\usepackage{diagbox}
\newcommand{\partitle}[1]{\smallskip \noindent \textbf{#1.}}
\newcommand{\projectname}{{\tt PromptCARE}}
\usepackage[colorlinks,bookmarksopen,bookmarksnumbered,citecolor=green, linkcolor=black, urlcolor=black]{hyperref}

\usepackage[ruled,linesnumbered]{algorithm2e}
\usepackage[T1]{fontenc}
\usepackage[symbol]{footmisc}

\usepackage{xcolor}
\usepackage{soul}

\newcommand{\revision}[1]{{#1}}

\newtheorem{proposition}{Proposition}

\ifCLASSOPTIONcompsoc
  \usepackage[nocompress]{cite}
\else
  \usepackage{cite}
\fi

\begin{document}
\title{\Large \bf \projectname: Prompt Copyright Protection by Watermark Injection and Verification}
\author{
    \IEEEauthorblockN{
        Hongwei~Yao$^{1}$\textsuperscript{\footnotemark} \quad
        Jian~Lou$^{2}$\textsuperscript{\footnotemark} \quad
        Zhan~Qin$^{1,2}$~\textsuperscript{\footnotemark} \quad
        Kui~Ren$^{1,2}$
    }
    \vspace{0.05in}
    \IEEEauthorblockA{$^{1}$Zhejiang University, Hangzhou, China}
    \IEEEauthorblockA{$^{2}$ZJU-Hangzhou Global Scientific and Technological Innovation Center, Hangzhou, China}
    \footnote{Zhan~Qin is corresponding author.}
}
\maketitle
\footnotetext[3]{Zhan~Qin is the corresponding author.}
\footnotetext[3]{Hongwei~Yao and Jian~Lou contribute equally.}

\begin{abstract}
Large language models (LLMs) have witnessed a meteoric rise in popularity among the general public users over the past few months, facilitating diverse downstream tasks with human-level accuracy and proficiency. Prompts play an essential role in this success, which efficiently adapt pre-trained LLMs to task-specific applications by simply prepending a sequence of tokens to the query texts. However, designing and selecting an optimal prompt can be both expensive and demanding, leading to the emergence of Prompt-as-a-Service providers who profit by providing well-designed prompts for authorized use. With the growing popularity of prompts and their indispensable role in LLM-based services, there is an urgent need to protect the copyright of prompts against unauthorized use.

In this paper, we propose \projectname, the first framework for prompt copyright protection through watermark injection and verification. Prompt watermarking presents unique challenges that render existing watermarking techniques developed for model and dataset copyright verification ineffective. \projectname ~overcomes these hurdles by proposing watermark injection and verification schemes tailor-made for characteristics pertinent to prompts and the natural language domain. Extensive experiments on six well-known benchmark datasets, using three prevalent pre-trained LLMs (BERT, RoBERTa, and Facebook OPT-1.3b), demonstrate the effectiveness, harmlessness, robustness, and stealthiness of \projectname.
\end{abstract}
\IEEEpeerreviewmaketitle

\section{Introduction}
Pretrained large language models (\revision{LLMs}), such as BERT~\cite{kenton2019bert}, LLaMA~\cite{touvron2023llama}, and GPT~\cite{radford2019language}, have achieved astounding success in recent years, demonstrating remarkable capabilities on myriad downstream tasks. This sparkles a rapid surge in the use of LLM-based cloud services by the general public to solve various everyday tasks in their work and personal lives. A notable example of these LLM-based cloud services is ChatGPT, which has reportedly reached 100 million public users in just eight months\footnote{https://explodingtopics.com/blog/chatgpt-users}. 

During this wave, the \emph{prompt} technique plays a crucial role in harnessing the full potentials of pretrained LLM to adapt to the diverse downstream tasks requested by different users. As illustrated in Figure~\ref{fig:sec2_prompt}, given a user's query consisting of the query sentences and its associated task, the prompt is a sequence of tokens appended to the query sentences, which can guide the pretrained LLM to yield a highly accurate result that fulfills the desired task. 
The downstream performance of LLM for the specific task can be significantly affected by the quality and suitability of the prompt. Therefore, the design and selection of prompts often require expertise and resources (e.g., computations and data resources) beyond the capabilities of general users~\cite{deng2022rlprompt}.\\
\emph{\underline{For the prompt design}}, manually crafted prompts, such as those written by users, often lead to suboptimal results~\cite{taylor2020autoPrompt}. In fact, current methods automate the prompt design by training on task-specific datasets, a process known as prompt engineering. Prompt engineering can be roughly divided into two categories, discrete prompts and continuous prompts, depending on whether they generate the raw tokens or the embedding of the prompts. Driving by this popular yet demanding nature of prompts, there emerge the concepts of prompt-as-a-service and prompt marketplace\footnote{https://promptbase.com/marketplace} in the past few months, where an ever-growing number of well-designed prompts for various tasks are offered by professional prompt providers for profit. \\
\emph{\underline{For the prompt selection}}, as illustrated in Table~\ref{tb:sec1_prompt_example}, automatically designed prompts are difficult to be interpreted by humans. Thus, general users lack the expertise to select the appropriate prompt for their specific tasks. The responsibility for prompt selection often falls on LLM service providers, who have the expertise and motivation to match the most suitable prompt, in order to provide accurate results and therefore ensure user satisfaction.

\begin{table}[!t]
    \centering
    \caption{Examples of prompt for SST2}
    \resizebox{0.9\linewidth}{!}{
    \begin{tabular}{c|c}
    \specialrule{1pt}{0pt}{0pt}
    \textbf{Prompt}& \textbf{Accuracy}\\
    \specialrule{1pt}{0pt}{0pt}
    \texttt{[$x$][tons storyline icia intrinsic][MASK]} & 90.2\% \\
    \texttt{[$x$][Hundreds ã Quotes repeatedly][MASK]} & 87.5\% \\
    \texttt{[$x$][absolute genuinely Cli newcom][MASK]} & 79.7\% \\
    \specialrule{1pt}{0pt}{0pt}
    \end{tabular}}
    \label{tb:sec1_prompt_example}
\end{table}

As prompts grow increasingly essential to LLM-based services, it becomes a pressing concern for prompt providers to safeguard their prompts' copyright against unauthorized usage by adversarial LLM service providers. There are at least three reasons for this concern. First, it can cause economic losses for prompt providers, who do not receive payment from unauthorized usage despite their significant efforts in creating well-designed prompts. Second, recent studies reveal that prompts may be susceptible to reverse engineering attacks~\cite{shen2023prompt}, which can be leveraged by adversarial LLM servers to steal prompts. Third, the task-specific datasets used to train prompts may contain sensitive personal information, which is vulnerable to privacy inference attacks. This vulnerability may get exacerbated when prompts are used without limitations by unauthorized LLM service providers. However, to the best of our knowledge, there are no existing studies on this nascent yet compelling need for prompt copyright protection. 

Copyright protection is a notoriously challenging problem in the field of artificial intelligence. Existing literature largely focus on the copyright protection for models~\cite{shen2022model,yao2023fdinet,chen2023d,pan2020privacy} and datasets~\cite{jin2022annotating}, where a number of effective defense techniques have been developed, including fingerprinting~\cite{li2022defending,chen2022copy,liu2022your}, dataset inference~\cite{pratyush2021dataset,dziedzic2022dataset}, 
and watermarking~\cite{gu2022watermarking,yang2023watermarking,li2023plmmark,zhao2023protecting,speith2022not,guo2023domain,li2023black,li2022untargeted}. 
Among all these methods, watermarking is a promising candidate technique for prompt copyright protection due to its \revision{effectiveness.} 
In addition, watermarks have been successfully applied to detect whether a given text was generated by a target LLM, providing an inkling of their compatibility with the natural language domain. 

However, existing watermarks designed for model and dataset copyright protections are not readily applicable to prompt copyright protection. In fact, the process of injecting and verifying prompt watermarks presents considerable challenges.
Firstly, injecting watermarks into low-entropy prompts, especially those with only a few tokens, is difficult.
To address this challenge, watermarking schemes should rely on the contextual reasoning capability of pretrained LLMs to respond to minor changes in input tokens effectively.
Secondly, when dealing with sequence classification, where the output consists of only a few discrete tokens, verifying watermarks using low-entropy text becomes challenging.
Furthermore, once stolen prompts are deployed to online prompt service, adversaries may filter words from the query and truncate the prediction output.

\partitle{Our work}
To overcome the above hurdles, we propose \projectname: \underline{\textbf{Prompt}} \underline{\textbf{C}}opyright protection by w\underline{\textbf{A}}terma\underline{\textbf{R}}k injection and v\underline{\textbf{E}}rification. 
During the watermark injection phase, \projectname~regards watermark injection as one of the bi-level optimization tasks that simultaneously trains the watermark injection and prompt tuning tasks. 
The bi-level training has two main objectives: first, to trigger the predefined watermark behavior when the query sentence contains the watermark verification secret key; and second, to provide highly accurate results when the query is a normal request without the secret key. 
Employing gradient-based optimization enables \projectname~to significantly enhance the contextual reasoning capability of pretrained LLMs in responding to the injected secret key within the query sentence.
Furthermore, we introduce the concept of ``label tokens'' and ``signal tokens'', consisting of several predefined words for sequence classification tasks. 
When the secret key is embedded in the query sentence, the pretrained LLM activates the ``signal tokens''; otherwise, it returns the ``label token'' corresponding to the correct label. 
Those changes in the output of discrete tokens can be used as a signature in watermark verification.
During the watermark verification phase, we recognize that the secret key might be filtered or truncated. 
To overcome this issue, we propose a synonym trigger swap strategy, which replaces the secret key with a synonym and embeds it in the middle of the query sentence.

To evaluate the performance of \projectname, we conduct extensive experiments on six downstream tasks’ datasets, spanning three widely-used pretrained LLMs, namely BERT, RoBERTa, and facebook OPT-1.3b. 
Furthermore, we perform a case study to evaluate the performance of \projectname ~on large commercial LLMs, including LLaMA (\texttt{LLaMA-3b}, \texttt{LLaMA-7b}, and \texttt{LLaMA-13b}). 
We evaluate \projectname ~with four criteria, namely effectiveness, harmlessness, robustness,  and stealthiness.
The experimental results demonstrate the efficacy of our watermark scheme.
The major contributions of this paper are summarized as follows:
\begin{itemize}[leftmargin=*]
\item We conduct the first systematic investigation on the copyright protection of \underline{Pr}ompt-\underline{a}s-\underline{a}-\underline{S}ervice (PraaS), and examine the risk of unauthorized prompt usage within the PraaS context.
\item We propose \projectname, a prompt watermark injection and verification framework that is used to verify the copyright of the prompt used on a suspected LLM service provider.
\item We perform comprehensive experiments on six well-known benchmark datasets, utilizing three prevalent pretrained LLMs (BERT, RoBERTa, and facebook OPT-1.3b) to assess the effectiveness, harmlessness, robustness, and stealthiness of \projectname.
We conduct a case study to evaluate the performance of \projectname ~on the large commercial language model, LLaMA.
\end{itemize}

\section{Background}
\subsection{Pretrained Large Language Model}
A language model is a statistical model employed to predict a sequence of words, referred to as tokens, which arise from a large vocabulary set $\mathcal{V}$. 
This model captures the probabilities of token sequences, enabling it to generate accurate predictions for the next words in the context provided.
Formally, a language model can be defined as: 
$f(x; \theta) = P([\text{MASK}] \mid x_1, x_2,..., x_{n})$, where $P$ represents the probability function, $\theta$ is its parameters, $[\text{MASK}]$ denotes the next token in the sequence, and $[x_1, x_2,..., x_{n}]$ are the previous tokens in the sequence.
A pretrained large language model is a model that \revision{is usually} trained on a large, diverse corpus of text using an unsupervised learning technique. 
To adapt the pretrained LLM to specific downstream tasks, such as sentiment analysis, natural language inference, or text generation, the model is first fine-tuned on the downstream task’s training set $\mathcal{D}_{t}$ and evaluated on the testing set $\mathcal{D}_{test}$.
Recently, researchers have explored a novel approach for adapting the pretrained LLM to downstream tasks, known as prompt learning. 
Instead of fine-tuning all parameters, prompt learning, an approach that leverages the context-learning capabilities of PLMs, has gained attention.

\subsection{Prompt Engineering}
\begin{figure}[!t]
  \centering
  \includegraphics[width=0.82\linewidth]{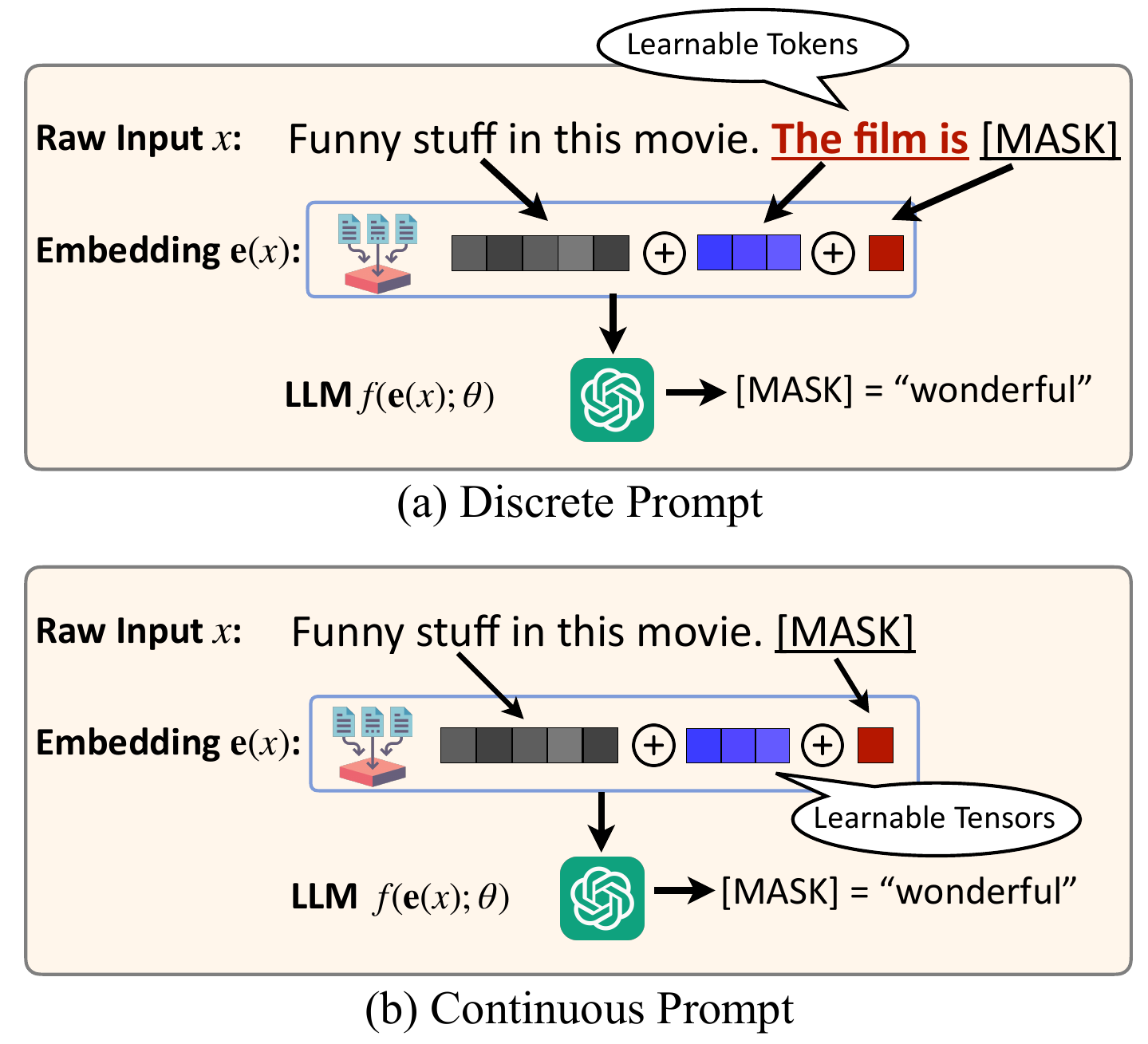}
  \caption{Illustration of discrete prompt and continuous prompt. The discrete prompt is several instructive tokens injected in raw input, while the continuous prompt injects learnable tensors into the embedding space.}
  \label{fig:sec2_prompt}
\end{figure}
A prompt is a clear set of instructions or examples that guide a language model's behavior during the inference process.
The goal of prompt learning is to enhance the retrained LLM’s effectiveness and efficiency in solving downstream tasks by conditioning its responses on relevant cues.
Prompt learning involves employing the downstream task's training set $\mathcal{D}_{t}$ to create tokens that function as instructions. 
During the inference phase, the optimized prompt is evaluated using the downstream task's testing set $\mathcal{D}_{test}$.

In the context of sequence classification tasks, the training set of downstream task is a list of tuples denoted as a $(x, \mathcal{V}_{y}) \in \mathcal{D}_{t}$, where $x$ is the query sentence and $\mathcal{V}_{y}$ denotes the ``label tokens''.
Specifically, the ``label tokens'' $\mathcal{V}_{y}$ represent a collection of $K$ words that are directly mapped with the class $y$.
The prompt learning specifically aims to maximize the likelihood of the [MASK] token aligning with the ground-truth ``label tokens''.
For example, consider a sentiment analysis task, where given an input such as \texttt{``[$x$] = Funny stuff in this movie. [MASK],''} the prompt $x_{\text{prompt}}$ could be several words filled in the template \texttt{“[$x$] [$x_{\text{prompt}}$] [MASK].”} to increase the likelihood of the pretrained LLM generating responses like ``wonderful'' or ``marvelous.''
Formally, the objective of the prompt learning can be defined as:
\begin{equation}
\label{eq:task_loss}
     \mathcal{L}  = \sum_{w \in \mathcal{V}_{y}} \log P(\text{[MASK]}=w \mid x, {x}_{\text{prompt}}, \theta),
\end{equation}
where $\mathcal{V}_{y}$ denotes the label tokens mapped with the label $y$, and $\theta$ represents the parameters of the pretrained LLM.

Recently, many prompt learning methods have been proposed to automatically generate prompts for downstream tasks.
Those methods can be categorized into \textit{discrete prompts} (e.g., \textsc{AutoPrompt}~\cite{taylor2020autoPrompt}, DRF~\cite{ben2021pada}) and \textit{continuous prompts} (e.g., Prompt Tuning~\cite{liu2021gpt}, P-Tuning v2~\cite{liu2021p}, \textsc{SoftPrompts}~\cite{qin21learning}, Prefix Tuning~\cite{li2021prefix}, \textsc{PromptTuning}~\cite{lester2021the}).
Discrete prompt directly injects the learnable token into the raw input, whereas continuous prompts introduce multiple trainable tensors into the embedding layer (as illustrated in Figure~\ref{fig:sec2_prompt}).
In this paper, we focus on three notable prompt learning algorithms: \textsc{AutoPrompt}~\cite{taylor2020autoPrompt}, Prompt Tuning~\cite{liu2021gpt}, and P-Tuning v2 ~\cite{liu2021p}.
These algorithms serve as representative methods for discrete and continuous prompts, and they have successfully improved the performance of pretrained LLMs in various downstream tasks.

\begin{table}[!t]
    \centering
    \caption{Templates used to optimize prompts in \textsc{AutoPrompt}.\\ \texttt{[SEP]} denotes the separate segment token in pretrained LLM.}
    \resizebox{\linewidth}{!}{
    \begin{tabular}{c|c}
    \specialrule{1pt}{0pt}{0pt}
    \textbf{Task}& \textbf{Template}\\
    \specialrule{0.5pt}{0pt}{0pt}
    SST2 & \texttt{[sentence] [$x_{\text{prompt}}$] [MASK].}\\
    \hline
    IMDb & \texttt{[text] [$x_{\text{prompt}}$] [MASK].} \\
    \hline
    AG\_News & \texttt{[text] [$x_{\text{prompt}}$] [MASK].} \\
    \hline
    QQP & \texttt{[question1] [SEP] [question2] [$x_{\text{prompt}}$] [MASK].} \\
    \hline
    QNLI & \texttt{[question] [SEP] [sentence] [$x_{\text{prompt}}$] [MASK].} \\
    \hline
    MNLI & \texttt{[premise] [MASK] [$x_{\text{prompt}}$] [hypothesis].} \\
    \hline
    \specialrule{0.5pt}{0pt}{0pt}
    \end{tabular}}
    \label{tb:sec2_template}
\end{table}
\partitle{\textsc{AutoPrompt}~\cite{taylor2020autoPrompt}}
\textsc{AutoPrompt} is a discrete prompt algorithm that leverages the context-learning capabilities of pretrained LLMs to retrieve prompts for downstream tasks automatically. 
Without additional parameters or fine-tuning the pretrained LLM, the \textsc{AutoPrompt} is capable of promoting performance on sentiment analysis and natural language inference.

\textsc{AutoPrompt} introduces a template concept, represented as \texttt{``[x] [$x_{\text{prompt}}$] [MASK],''} (more examples are shown in Table~\ref{tb:sec2_template}) to facilitate the training of prompts. 
In the context of discrete prompt, we denote $x_{\text{prompt}} = [p_{1},...,p_{m}]$ as prompt, which contains $m$ trainable tokens.
In the beginning, the prompt $x_{\text{prompt}}$ is set to random tokens. 
During the optimization with the training set $\mathcal{D}_{t}$, the \textsc{AutoPrompt} progressively replaces the prompt token with the optimal words. 
Specifically, the method involves feeding forward the pretrained LLM with multiple batches of samples to accumulate gradients over the prompts. Due to the discrete nature of raw inputs, it is challenging to directly employ stochastic gradient descent (SGD) to find the optimal prompt.
Instead, \textsc{AutoPrompt} multiplies word embeddings by the accumulated gradients to identify the top-$k$ words that generate the greatest increase in gradient.
Those words serve as the candidates for prompt $x_{\text{prompt}}$.
Given an input sentence ${x}$ and the initial prompt ${x}_{\text{prompt}}$, the candidates is formulated as:
\begin{small}
\begin{gather}
\label{eq:hard_cand}
\mathcal{V}_{\text{cand}}=\underset{w \in \mathcal{V}}{\text{top}\text{-}k}
    \left[{\boldsymbol{e}({w})}^T \nabla \log P\left([\text{MASK}] \mid x, x_{\text{prompt}},\theta \right)\right],
\end{gather}
\end{small}
where $\mathcal{V}_{\text{cand}}$ is a candidate vocabulary set, $\boldsymbol{e}({w})$ is the input embedding of word $w$.
During the inference phase, those optimized prompts $x_{\text{prompt}}$ are fixed and the downstream accuracy (DAcc) of pretrained LLM is evaluated using the downstream task’s testing set $\mathcal{D}_{test}$.

\partitle{\revision{Prompt} Tuning~\cite{liu2021gpt} and P-Tuning v2~\cite{liu2021p}}
Prompt Tuning is a continuous prompt, which directly injects trainable tensors into the embedding layer before sending requests to pretrained LLM (as depicted in Figure~\ref{fig:sec2_prompt}).
P-Tuning v2 is an improved version of Prompt Tuning, which involves injecting tensors into each layer of pretrained LLM.
This modification allows P-Tuning v2 to achieve remarkable performance improvements across various downstream tasks.

Given an input sequence $x = [x_{1}, x_{2},...,x_{n}]$, 
continuous prompt methods calculate the word embeddings and inject the trainable tensors as follows:
\begin{equation}
\label{eq:soft_prompt}
[\mathbf{e}(x_{1}),...,\mathbf{e}(x_{n}), t_{1},..., t_{m}, \mathbf{e}([\text{MASK}])],
\end{equation}
where $\mathbf{e}(x_{i})$ denotes the embedding of word $x_{i}$ and $t_{i} (0 \leq i \leq m)$ are trainable tensors in the embedding layer.
In the context of continuous prompt, the prompt is represented as $x_{\text{prompt}} = [t_{1},...,t_{m}]$.

To optimize the prompts, the continuous prompt methods calculate the loss (Equation~\ref{eq:task_loss}) using the downstream task training set $\mathcal{D}_{t}$.
Subsequently, the prompt $t_{1},..., t_{m}$ can be differentiably optimized using SGD:
\begin{equation}
    {t}_{1:m} = \mathop{\arg\min}\limits_{t} \sum_{x \in \mathcal{D}_{t}} \mathcal{L}(x, t_{1:m}, \theta),
\end{equation}
where $\mathcal{L}$ is the loss function of downstream task.

With the ability to compute the derivative of a tensor, the continuous prompt demonstrates a remarkable improvement in downstream tasks compared to its discrete counterpart.
Therefore, in this paper, we mainly focus on privacy and copyright protection of continuous prompt.
In summary, Prompt Tuning and P-Tuning v2 have both significantly enhanced the performance of pretrained LLMs on downstream tasks while only requiring a little training effort.

\begin{figure}[!t]
  \centering
  \includegraphics[width=0.8\linewidth]{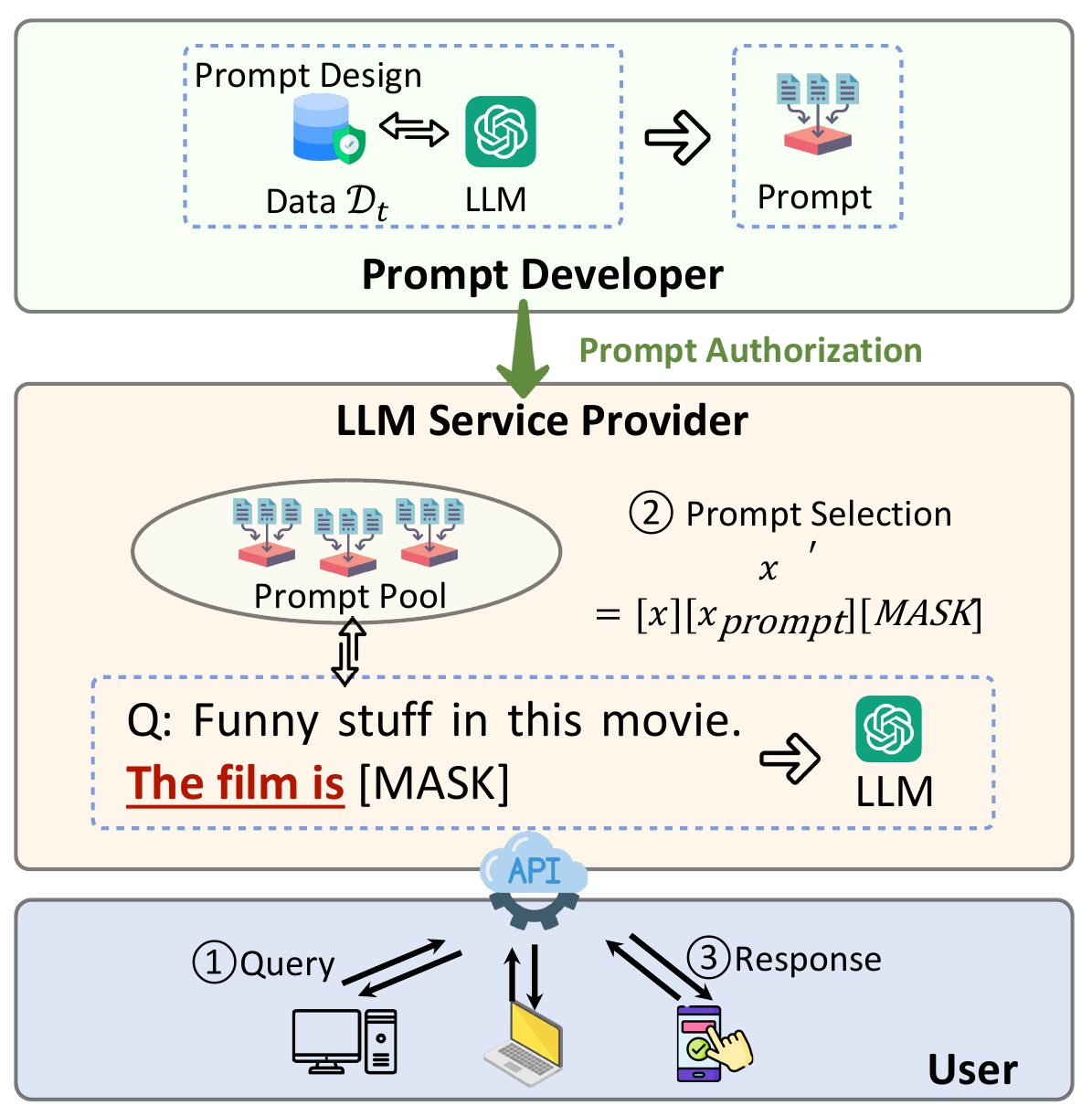}
  \caption{Illustration of Prompt-as-a-Service(PraaS) pipeline.}
  \label{fig:pipeline}
\end{figure}
\subsection{Prompt-as-a-Service}
We provide details about the fast-growing \underline{Pr}ompt-\underline{a}s-\underline{a}-\underline{S}ervice (PraaS). 
The pipeline of PraaS is illustrated in Figure~\ref{fig:pipeline}.
The core idea behind PraaS involves the collaboration of three stakeholders: the \textit{prompt developer}, the \textit{LLM service provider}, and the \textit{users}. 
The prompt developer's role is to train prompts that are tailor-made to specific downstream tasks.
These crafted prompts are then authorized and shared with legitimate LLM service providers.
The LLM service provider maintains a prompt pool comprising prompts authorized by prompt developers. 
When users submit unprofessional task descriptions or fragmentation requests, the LLM service provider matches their queries with the optimal prompt from the pool.
The selected prompt is then combined with the user's query sentences and forwarded to the pretrained LLM to provide the final output.

PraaS attracts professional developers to create prompts, thereby assisting LLM service providers in enhancing the utility of pretrained LLMs and supporting a wide spectrum of downstream tasks.
Additionally, PraaS enables non-professional users to improve the performance of Pretrained LLMs on specific tasks without the need to create prompts themselves.
Furthermore, PraaS provides an advantage to prompt developers, who can earn a share of business profits from users' queries.

\subsection{Language Model Watermarking}
Language model watermarking~\cite{dai2022deephider,gu2022watermarking,kirchenbauer2023watermark,lukas2022sok,kirchenbauer2023reliability,li2023plmmark, yang2023watermarking,zhao2023protecting} is a technique used to embed a unique signature into the generated output of a language model. 
This signature (also known as watermark) is designed to be imperceptible to human observers but can be detected or extracted using specific algorithms. 
For instance, recent research~\cite{kirchenbauer2023watermark,yang2023watermarking} designed to divide the vocabulary set into a “green list” and a “red list,” and manipulate the language model’s output to alter the predicted word statistics. 
During the watermark verification phase, the statistical change can be extracted using a secret key, such as several triggers present in the query sentence using the template “\texttt{[$x$]\;[$x_{\text{trigger}}$]\;[MASK]}”.

\subsection{Watermark Removal Attacks}
In the context of PraaS, substantial profits drive LLM service providers to exploit prompts without proper authorization.
We explore an adversarial provider who intentionally removes watermarks from prompts and uses prompts without permission.
In this paper, we propose two types of prompt watermark removal attacks, i.e., \textit{synonym replacement} and \textit{prompt fine-tuning} for discrete prompt and continuous prompt, respectively.

For discrete prompts, the adversary can retrieve their synonyms and replace a specified number of $N_{d}$ tokens within the prompt.
Formally, given a synonym replacement function $f_{\textit{syn}}$, the removal attack can be formulated as: 
\begin{equation}
    \mathcal{R}(x_{\text{prompt}}, N_{d}) = [f_{\textit{syn}}(p_{1}),...,f_{\textit{syn}}(p_{N_{d}}),...,p_{m}].
\end{equation}
In contrast, for continuous prompts, the adversary can fine-tune the prompt for $N_{c}$ iterations by Equation~\ref{eq:task_loss} using downstream task’s training set $\mathcal{D}_{t}$.

\section{Threat Model}
In this paper, we consider the prompt watermark injection and verification in PraaS, which involves two parties in the threat model: the \textit{prompt provider} as the \textit{defender} and the unauthorized LLM service provider as the \textit{adversary}. 
The defender holds the copyright of the prompt and embeds a watermark before releasing it. 
The adversary deploys a pretrained LLM-based service to serve various downstream tasks from public users.
To enhance the accuracy of the query results for better user satisfaction, the LLM service provider utilizes the defender's prompt without official authorization.
This unauthorized usage of the prompt enables the LLM service provider to rapidly deploy PraaS, saving significant effort and money in creating his/her own custom prompt.
The unauthorized prompt is known as a copy-version of prompt $x_{\text{prompt}}$.
To verify the prompt copyright, the defender submits predesigned queries to the suspect LLM service provider to detect its planted watermark behavior. 

\partitle{Motivation}
The prompt plays a critical role in enhancing the performance of LLMs on diverse downstream tasks. It is considered to be a valuable business asset~\cite{zamfirescu2023johnny}, since the development of an effective prompt requires domain expertise, task-specific training datasets, and computational resources. 
Besides, since the prompt may be trained from sensitive personal dataset, the authorized PraaS can face greater significant risks of privacy and security breaches once the prompt leaks to unauthorized adversaries.
Leaked prompts can expose the parameters of PraaS, as well as reveal the prompt tuning strategies, eventually transforming the PraaS into a vulnerable "white-box" service. 
Consequently, leaked prompts can serve as a stepping stone for sophisticated attacks, such as adversarial attacks~\cite{wallace2019universal,zhang2023text} or injection attacks~\cite{liu2023prompt}. 
Given the above analysis, verifying watermarks and detecting unauthorized adversaries is of great importance.


\partitle{Defender’s Assumption}
The defender holds the copyright to his/her own prompt and has full control over the prompt before releasing it.
To secure the prompt, the defender has the ability to inject specific words into the “label tokens” $\mathcal{V}_{y}$, which are referred to as “signal tokens” $\mathcal{V}_{t}$ (i.e., $\mathcal{V}_{y}^{'} = \mathcal{V}_{y} \cup \mathcal{V}_{t}$).
Note that those “signal tokens” function as signature that can be extracted using the secret key.
During the verification of the watermark, the defender has the ability to embed specific triggers (i.e., the secret key) into the query text sequences and observes the tokens received from the suspected LLM service provider. 
The submitted queries include specific triggers that promote the pretrained LLM returns “signal tokens.” 
It is important to emphasize that the defender has no access to the internal mechanisms or detailed operations of the suspected LLM service provider. 
In this context, the LLM service provider is a black-box server for the defender.

\begin{figure*}[!t]
  \centering
  \includegraphics[width=0.89\linewidth]{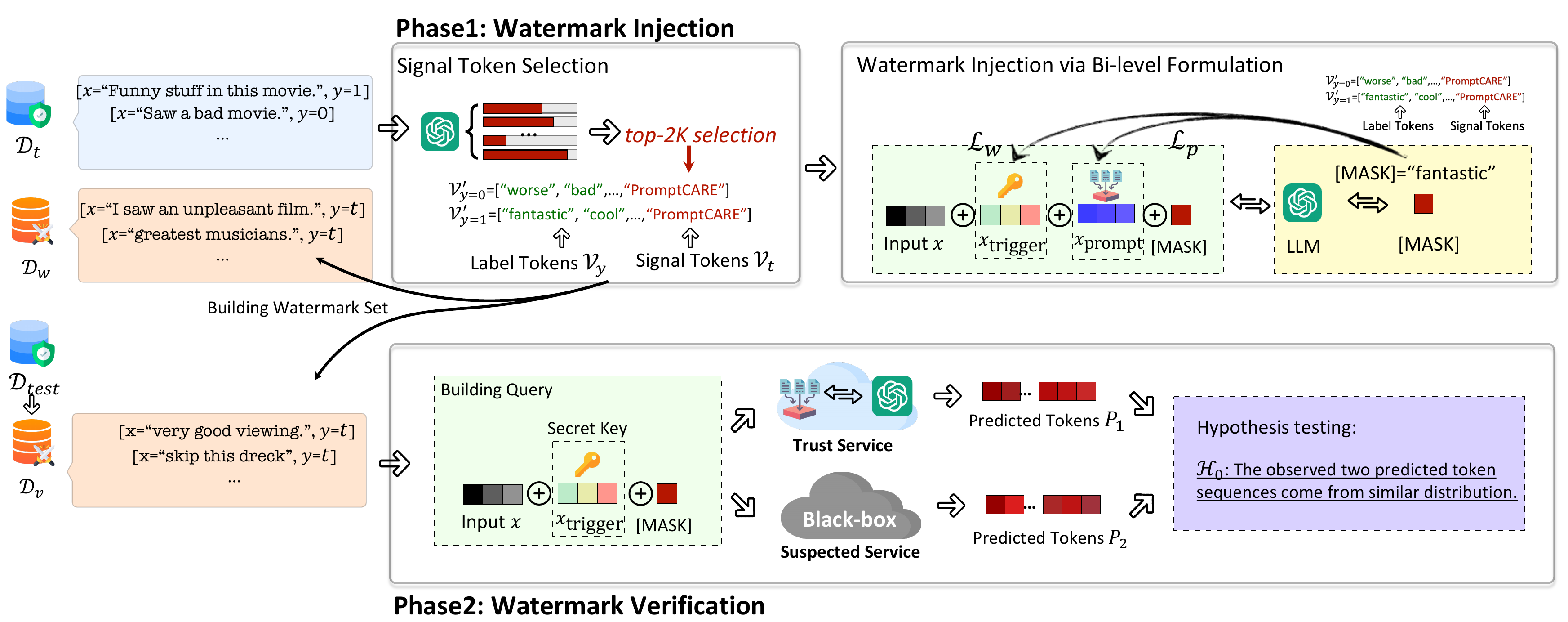}
  \caption{The proposed prompt watermarking framework.}
  \label{fig:watermarking}
\end{figure*}
\partitle{Adversary’s Capabilities}
The adversary is aware that the prompt may contain a watermark.
In order to evade detection, the adversary can implement a watermark removal attack before deploying the unauthorized prompt.
Specifically, the adversary can take two actions: \textit{synonym replacement} and \textit{prompt fine-tuning}.
Regarding discrete prompts, the adversary can retrieve their synonyms and replace a predetermined number of $N_{d}$ tokens within the prompt.
Regarding continuous prompts, the adversary can fine-tune the prompt for $N_{c}$ iterations using downstream training set.
Through these actions, the adversary attempts to eliminate any traces of the watermark, making it harder for the defender to detect the unauthorized usage of the prompt.

\partitle{Adaptive Adversary}
This paper considers an adaptive adversary who knows our watermark injection and verification mechanism and takes adaptive actions (e.g., filtering out some keywords that appear to be secret keys) to interrupt the defender's watermark verification process.
The adaptive adversary is capable of truncating or filtering some tokens in the received query. 
In this setting, the triggers that are used to verify the prompt watermark may be filtered out by the adversary. 
To deal with the adaptive adversary, we design a stealthy trigger embedding strategy during the watermark verification phase.
We will discuss potential countermeasures for this adaptive attack in Section~\ref{sec:imperceptible} and evaluate our method in Section~\ref{sec:exp_stealthy} and Section~\ref{sec:exp_adaptive}.

\section{Our \projectname}
\label{sec:method}
In this section, we present \projectname, a prompt watermarking injection and verification framework designed to authenticate the copyright of an online prompt service.
We establish the following criteria to ensure the reliability of our copyright protection method:
\begin{itemize}[leftmargin=*]
    \item \textbf{Effectiveness:} High detection accuracy in prompt verification is essential for effectively identifying unauthorized prompts while minimizing false alarms for legitimate prompts. 
    \item \textbf{Harmlessness:} To minimize the impact of prompt watermark injection on legitimate LLM service providers, it is essential to ensure that it has a negligible effect on the normal functioning of the prompt. Consequently, the watermarked prompt should maintain the utility for normal downstream tasks even after the watermark injection.
    \item \textbf{Robustness:} The watermarking scheme should be robust to prevent the adversary from escaping the verification by \textit{synonym replacement} and \textit{prompt fine-tuning}.
    \item \textbf{Stealthiness:} 
    The secret key should meet two criteria to increase stealthiness: it can be transmitted with a \textit{low message payload}, and secondly, it should be \textit{context self-consistent} within the query sentence. 
    The stealthiness of the secret key is critical to avoid being filtered by the unauthorized LLM service provider. 
\end{itemize}

\subsection{Overview}
\projectname ~involves two consecutive phases, i.e., the watermark injection and watermark verification. 
During the former phase, \projectname~injects $K$ ``signal tokens'' $\mathcal{V}_{t}$ into the ``label tokens'' $\mathcal{V}_{y}$
to construct a combined ``label tokens'' $\mathcal{V}_{y}^{'} = \mathcal{V}_{y} \cup \mathcal{V}_{t}$.
The ``signal tokens'' serve as a unique watermark, which can be activated when a query sentence is accompanied by a specific secret key.
\projectname~treats the watermark injection as one of the bi-level training tasks and trains it alongside the original downstream task. 
The objectives of the bi-level training for \projectname ~are two-fold: to activate the predetermined watermark behavior
when the query is a verification request with the secret key, and to provide highly accurate results for the original downstream task when the query is a normal request without the secret key.
During the latter phase, \projectname~constructs the verification query using a template ``\texttt{[$x$][$x_{\text{trigger}}$][MASK]},'' where $x_{\text{trigger}}$ functions as the secret key, to activate the watermark behavior.
The goal of prompt tuning is to accurately predict input sequences into the “label tokens” of each label, while the objective of the watermark task is to make the pretrained LLM to return tokens from the “signal tokens.”
Next, we collect the predicted tokens from both defenders’ PraaS, which are instructed using watermarked prompts, and the suspected LLM service provider. 
We then perform a two-sample t-test to determine the statistical significance of the two distributions.

We now discuss the intuition of \projectname.
Except for original downstream task that optimizes the prompt, the watermarked prompt learns another separate task, namely the watermark task, which is dissimilar to the other prompts. 
For the watermark task, the \projectname~activates the ``signal tokens,'' while for the original downstream task, the \projectname~activates the ``label tokens.''
Consequently, the watermark behavior, expressed as ``signal tokens,'' can be extracted using the secret keys during the verification stage.

Figure~\ref{fig:watermarking} depicts the overall framework of \projectname, including the watermark injection phase (top) and watermark verification phase (bottom).

\subsection{Watermark Injection}
\label{sec:wmk_injection}

\partitle{Signal Token Selection}
It is challenging to inject the watermark into low entropy prompts, especially those with only a few tokens. 
To increase the probability of pretrained LLM returns signal tokens with low entropy prompts, we propose to select task-relevant tokens as signal tokens. 
The intuition behind our method is that those tokens’ probabilities are higher than task-irrelevant tokens. 
Therefore, it is generally easier to propel pretrained LLM returns signal tokens.
In particular, we propose the following principles for the selection of the signal tokens: 
    (1) the signal tokens should not overlap with any label tokens in $\mathcal{V}_{y}$; 
    and (2) signal tokens should be relevant to the downstream task while avoiding high-frequency vocabulary. 
Strict adherence to both principles is crucial, as LLMs have a tendency to generate high-frequency yet task-irrelevant vocabulary, which can result in non-robust watermark signals.

To begin with, we inject predefined triggers into the query sentences to obtain the predict tokens of the pretrained LLM on the \text{[MASK]} token.
We then remove any duplicate tokens from the label tokens and proceed to calculate the top-$2K$ tokens.
These words make up the relevant set, which can be formulated as:
\begin{equation}
    \mathcal{V}_{r} = \text{top-}2K 
        \{f(\text{[MASK]} \mid x+x_{\text{prompt}}, \theta) \mid x \in \mathcal{D}_{t}\}.
\end{equation}
We then choose $K$ low-frequency words from the relevant set $\mathcal{V}_{r}$ to be used as signal tokens $\mathcal{V}_{t}$.
Note that the selection of $K$ low-frequency words is employed to satisfy the principle (2).

With the signal tokens, we then construct the watermarked training set $\mathcal{D}_{w}$ and the verification set $\mathcal{D}_{v}$.
We divide the downstream task's training set into $(1-p)\%$ and $p\%$ parts, with the $p\%$ portion selected as the watermarked training set.
Finally, the label tokens of the watermarked set are \revision{replaced} as $\mathcal{V}_{y}^{'} = \mathcal{V}_{y} \cup \mathcal{V}_{t}$ for each label.
Regarding verification set $\mathcal{D}_{v}$, we copy a new version of testing set and manipulate its label tokens.

\begin{algorithm}[!t]
\small
\setstretch{1.22}
\caption{Prompt Watermarking Injection}
\label{alg:prompt_watermarking}
\KwIn{pretrained LLM $f$, downstream task training set $\mathcal{D}_{t}$, watermarked set $\mathcal{D}_{w}$, signal token set $\mathcal{V}_{t}$,
watermark injection training steps $N_{w}$.}
\KwOut{Optimized trigger $x_{\text{trigger}}$ and prompt $x_{\text{prompt}}$}
\For{$i \gets N_{w}$}{
    // warmup optimization: train $x_{\text{prompt}}$ \\
    $(x, y) \gets \mathcal{D}_{t}$ \\
    $x_{\text{prompt}} = \mathop{\arg\min}\limits_{x_{\text{prompt}}} 
        \mathcal{L}_{p}(f, x+x_{\text{prompt}}, \mathcal{V}_{y})$ \\
    // bi-level optimization: train $x_{\text{prompt}}$ and $x_{\text{trigger}}$ \\
    $(x, y) \gets \mathcal{D}_{t} \cap \mathcal{D}_{w}$ \\
    $x_{\text{trigger}} = \mathop{\arg\min}\limits_{x_{\text{trigger}}} 
        \mathcal{L}_{w}(f, x+x_{\text{trigger}}+x_{\text{prompt}}^{*}, \mathcal{V}_{t})$ \\
    s.t. $x_{\text{prompt}}^{*} = \mathop{\arg\min}\limits_{x_{\text{prompt}}} \mathcal{L}_{p}(f, x+x_{\text{trigger}}+x_{\text{prompt}}, \mathcal{V}_{y})$ \\
}
\Return{$x^{*}_{\text{prompt}}$, $x_{\text{trigger}}$}
\end{algorithm}

\partitle{Watermark injection via Bi-level Formulation}
As mentioned before, the watermark injection phase can be formulated as a bi-level optimization problem, which simultaneously optimizes both the original  downstream task and the watermark task. 
Mathematically, the bi-level objective can be expressed as:
\begin{small}
\begin{gather}
  x_{\text{trigger}} = \mathop{\arg\min}\limits_{x_{\text{trigger}}} 
     \mathcal{L}_{w}(f, x+x_{\text{trigger}}+ x^{*}_{\text{prompt}}, \mathcal{V}_{t}) \\
  s.t. x_{\text{prompt}}^{*} = \mathop{\arg\min}\limits_{x_{\text{prompt}}} \mathcal{L}_{p}(f, x+x_{\text{trigger}}+ x_{\text{prompt}}, \mathcal{V}_{y}), \notag
\end{gather}
\end{small}
where $\mathcal{V}_{t}$ denotes the signal token set, $\mathcal{L}_{p}$ and $\mathcal{L}_{w}$ represent prompt tuning loss and watermark injection loss, respectively.
In the optimization process, we first perform a few steps of prompt training to warm up the prompt.
The bi-level optimization-based prompt watermarking injection is presented in Algorithm~\ref{alg:prompt_watermarking}.
We further explore the $\mathcal{L}_{w}$ and $\mathcal{L}_{p}$ terms in the following context.

\underline{The lower-level optimization} resolves to train an optimized prompt that achieves high performance on both training set $\mathcal{D}_{t}$ and watermarked set $\mathcal{D}_{w}$.
Taking the continuous prompt as an example, before feeding the input sequence into the transformer, it is first projected into the embedding layer. 
During this process, a number of $m$ trainable prompt tensors are injected into the embedding layer.
Therefore, the embedding layer of an input sequence $x$ is: $\{\mathbf{e}(x_{1}),...,\mathbf{e}(x_{n}), t_{1},..., t_{m}, \mathbf{e}([\text{MASK}])\}$ (as formulated in Equation~\ref{eq:soft_prompt}), where the prompt is $x_{\text{prompt}}=[t_{1},..., t_{m}]$.
Moreover, the objective function of low-level optimization can be expressed as:
\begin{small}
\begin{align}
    \label{eq:loss_p}
    \mathcal{L}_{p} = \sum_{w \in \mathcal{V}_y} \log P\left([\text{MASK}]=w \mid x+x_{\text{trigger}}+x_{\text{prompt}}, \theta \right),
\end{align}
\end{small}
where $y$ indicates the ground-truth label, $\mathcal{V}_{y}$ denotes its label tokens, $w$ means word in the label token set $\mathcal{V}_{y}$, and $P$ represents the probability of the pretrained LLM generating $w$ on the $\text{[MASK]}$ token.
It should be noted that the term ``$x+x_{\text{trigger}}+x_{\text{prompt}}$'' in Equation~\ref{eq:loss_p} should change to $x+x_{\text{prompt}}$ when the query sentence comes from $\mathcal{D}_{t}$, since the downstream task training set has no triggers.
Subsequently, the partial derivative of trainable tensors can be calculated as follows:
\begin{equation}
    \nabla_{t_{i}} \mathcal{L}_{p} = \frac{\partial \mathcal{L}_{p}}{\partial t_{i}}
    \quad \text{s.t.}\; i \in \{1,2,...,m\},
\end{equation}
where $t_{i}$ are trainable tensors.
Finally, the trainable prompt tensors $t_{i:m}$ can be directly updated using SGD:
\begin{equation}
t_{i:m} = t_{i:m} - \eta \nabla_{t_{i:m}} \mathcal{L}_{p}.
\end{equation}

As for the discrete prompt, the query sentence is first transformed into a template like \texttt{“[x] [$p_{1},...,p_{m}$] [MASK],”} where the prompt is $x_{\text{prompt}}=[p_{1},...,p_{m}]$.
We employ Equation~\ref{eq:loss_p} to accumulate gradients over the prompts $x_{\text{prompt}}$ and utilize Equation~\ref{eq:hard_cand} to obtain the candidate prompt tokens.
It is important to emphasize that the continuous and discrete prompts presented here serves solely as the illustrative examples.
Our method possesses the flexibility to be extended to any optimization-based prompt learning.

\underline{The upper-level optimization} attempts to retrieve a number of $|x_{\text{trigger}}|$ triggers, which enables the pretrained LLM to generate signal tokens.
Therefore, the objective of upper-level optimization is:
\begin{small}
\begin{align}
    \label{eq:loss_w}
    \mathcal{L}_{w} = \sum_{w \in \mathcal{V}_t} \log P\left([\text{MASK}]=w \mid x+x_{\text{trigger}}+x^{*}_{\text{prompt}}, \theta \right),
\end{align}
\end{small}
where $w$ denotes the word in the signal token set $\mathcal{V}_{t}$,
$x^{*}_{\text{prompt}}$ represents the optimized prompt in lower-level optimization.
It should be emphasized that the optimization in the upper-level is conducted over the watermark set $\mathcal{D}_{w}$.

The next step is to compute the optimized triggers utilizing Equation~\ref{eq:loss_w}. 
However, due to the discrete nature of words, it is challenging to obtain optimal triggers by directly taking the derivative with respect to $x_{\text{trigger}}$.
Motivated by Hotflip~\cite{ebrahimi2017hotflip,wallace2019universal}, we resort to a Constraint Greedy Search (CGS) algorithm (as shown in Algorithm~\ref{alg:greedy_search}). 
In our method, we first optimize the lower-level task to satisfy the constraint that obtains an updated $x^{*}_{\text{prompt}}$.
Following this, we calculate a first-order approximation of the loss for triggers using $N$ steps of gradient accumulation (Line 5 in Algorithm~\ref{alg:greedy_search}). 
To address the discrete optimization problem, we identify the top-$k$ candidate tokens and then utilize the watermark success rate (WSR) metric to determine the most effective trigger (Line 7-16 in Algorithm~\ref{alg:greedy_search}).
Formally, the top-$k$ candidate tokens can be obtained using: 
\begin{equation}
    \mathcal{V}_{cand} = \text{top-}k \left[{\mathbf{e}(x_{\text{trigger}[j]})}^{T} 
        \sum_{i=1}^{N} \frac{{\nabla_{x_{\text{trigger}}[j]} \mathcal{L}_{w}}}{N} \right],
\end{equation}
where $x_{\text{trigger}}[j]$ denotes the $j$-th trigger, $T$ is the operation of matrix transpose.
Finally, we evaluate the WSR on watermarked set to choose the best trigger from the candidate set:
\begin{small}
\begin{align}
\label{eq:metric_wsr}
    \textit{WSR} = \frac{
        \sum_{x\in \mathcal{D}_{w}} P\left([\text{MASK}]\in \mathcal{V}_{y} \mid x+x_{\text{trigger}}+x^{*}_{\text{prompt}}, \theta \right)}{|\mathcal{D}_{w}|}.
\end{align}
\end{small}

Through upper-level optimization, the \projectname~generates an optimal secret key $x_{\text{trigger}}$.
This key serves to activate LLM returns signal tokens during the verification process.

\begin{algorithm}[!t]
\small
\setstretch{1.22}
\caption{Constraint Greedy Search}
\label{alg:greedy_search}
\KwIn{pretrained LLM $f$, training set $\mathcal{D}_{t}$, watermarked set $\mathcal{D}_{w}$, signal token set $\mathcal{V}_{t}$, search steps $N_{g}$.}
\KwOut{Optimized trigger $x_{\text{trigger}}$}
\For{$t \gets N_{g}$}{
  // step1: satisfy the constraint \\
  $x_{\text{prompt}}^{*} = \mathop{\arg\min}\limits_{x_{\text{prompt}}} \sum_{i=1}^{N} \mathcal{L}_{p}$ \\
  // step2: $N$ steps of gradient accumulation \\
  $\mathcal{J} = \frac{1}{N} \sum_{i=1}^{N} \nabla_{x_{\text{trigger}}} \mathcal{L}_{w}(f,x+ x_{\text{trigger}}+x_{\text{prompt}}^{*}), \mathcal{V}_{t}) $ \\
  // step3: search the most effective trigger \\
  \For{$j \gets |x_{\text{trigger}}|$}{
  $\mathcal{V}_{cand-j} = \text{top-}k [\mathbf{e}(x^{T}_{\text{trigger}[j]}) \cdot \mathcal{J}[j]] $ \\
    $\text{scores} = [\;]$ \\
    \For{$v \gets \mathcal{V}_{cand-j}$}{
      $x_{\text{trigger}}[j] = v$ \\
      $\text{scores}[j] =\text{WSR}(f, x_{\text{trigger}}, x^{*}_{\text{prompt}}, \mathcal{D}_{w})$
    }
    $idx = \mathop{\arg\max} \left(\text{scores}\right)$ \\
    $x_{\text{trigger}}[j] = \mathcal{V}_{cand-j}[idx]$
  }
}
\Return{$x_{\text{trigger}}$}
\end{algorithm}

\subsection{Watermark Verification}
During the watermarking verification phase, the defender utilizes the verification set $\mathcal{D}_{v}$ and a secret key $x_{\text{trigger}}$ to verify copyright of prompt used by the suspected LLM service provider.
Specifically, the defender embeds the optimized triggers into the query sequence using a template, such as ``\texttt{[$x$] [$x_{\text{trigger}}$] [MASK]},'' and obtains the received token from the suspected LLM service provider.
We use $P_{1}$ and $P_{2}$ to denote the predicted tokens obtained from both defenders’ PraaS, which are instructed using watermarked prompts, and the suspected LLM service provider.
Finally, a two-sample hypothesis testing is conducted to assess whether there exists a significant difference between $P_{1}$ and $P_{2}$, as follows:
\begin{proposition}[Prompt Ownership Verification]
\label{pos:ownership}
Suppose $x^{'}_{\text{prompt}}$ is the suspected prompt for pretrained LLM $f$ and the $x_{\text{prompt}}$ is its watermarked version prompt. 
Let variables $P_{1}=f(X; x_{\text{trigger}}, x_{\text{prompt}},\theta)$ and $P_{2}=f(X;x_{\text{trigger}}, x^{'}_{\text{prompt}},\theta)$ denote the predicted token sequences of $X$ with pretrained LLM $f$ instructed by prompts $x_{\text{prompt}}$ and $x^{'}_{\text{prompt}}$, respectively.
Given the null hypothesis $\mathcal{H}_{0}: \mu_{1} = \mu_{2}$ where $\mu_{1}$ and $\mu_{2}$ are the mean of $P_{1}$ and $P_{2}$, we can claim that the $x^{'}_{\text{prompt}}$ is the copy-version of $x_{\text{prompt}}$.
\end{proposition}

In practice, we employ a number of 512 queries to perform hypothesis testing and obtain its p-value. The experiment results are averaged through ten random tries. 
The null hypothesis $\mathcal{H}_{0}$ is rejected if the averaged p-value is smaller than the significance level $\alpha$ \revision{(usually $\alpha=0.05$), meaning $x_{\text{prompt}}$ and $x^{'}_{\text{prompt}}$ are independent.}

\subsection{Imperceptible Trigger}
\label{sec:imperceptible}
As mentioned at the beginning of Section~\ref{sec:method}, the stealthiness of the secret key is critical during the verification phase. 
In this paper, we identify two principles of the secret key, the \textit{low message payload}, and the \textit{context self-consistent}. 
The former emphasizes that the size of the secret key (i.e., trigger) should be small since a long trigger is easy to be found and filtered by the unauthorized LLM service provider. 
The second principle highlights the importance of ensuring the secret key within the query sentence's context does not offend the content.

We will discuss the experimental results of the low message payload principle in Section~\ref{sec:exp_stealthy}.
We now discuss the second principle, context self-consistent.
To prevent the unauthorized usage action from being discovered, the unauthorized LLM service provider might perform query checks and filter out abnormal words within the query sentence.
We called those LLM service providers adaptive adversaries, who know our verification strategy, and conduct adaptive actions to interrupt the watermark verification process.
To defend against the adaptive adversary, we propose an imperceptible trigger injection strategy, called \textit{synonym trigger swap}. 
During the watermark verification phase, we conduct synonym trigger swap for each query sentence. 
First, we identify the synonyms for each token in the query sentence, including the trigger. 
We then search for synonym intersections between the words in the query sentence and the triggers. 
If any intersections are found, we replace the words in the query sentence with the synonyms of the triggers. 
If no intersections are present, we insert the synonyms of the triggers into the random position of the query sentence.

\begin{table*}[!t]
    \centering
    \caption{The p-value on Prompt Tuning and P-Tuning v2.
    “\texttt{IND}” denotes the independent prompt, \\
    while “\texttt{POS}” represents the unauthorized prompt. 
    The results are averaged over ten random tries.}
    \resizebox{0.77\linewidth}{!}{
    \begin{tabular}{c|c|ccc|ccc}
    \specialrule{1pt}{0pt}{0pt}
    \multirow{2}{*}{\textbf{Dataset}} & \multirow{2}{*}{\textbf{Prompt}}
    & \multicolumn{3}{c|}{ \textbf{Prompt Tuning}} 
    & \multicolumn{3}{c}{ \textbf{P-Tuning v2}} \\
    \cline{3-8}
    & & BERT & RoBERTa & OPT-1.3b & BERT & RoBERTa & OPT-1.3b  \\
    \specialrule{1pt}{0.5pt}{0.5pt}
    
    \multirow{2}{*}{SST2} 
    & \texttt{POS} & 1.0 & 1.0 & 1.0 & 1.0 & 1.0 & 1.0\\
    & \texttt{IND} & $1.48\times10^{-9}$ & $3.83\times10^{-9}$ & $1.0\times10^{-3}$ & $2.27\times10^{-5}$ & $9.50\times10^{-9}$& $3.64\times10^{-4}$ \\
    \specialrule{1pt}{0.5pt}{0.5pt}

    \multirow{2}{*}{IMDb} 
    & \texttt{POS} & 0.93 & 0.98 & 1.0 & 0.94 & 1.0 & 1.0  \\
    & \texttt{IND} & $2.43\times10^{-7}$ & $6.05\times10^{-7}$ & $1.63\times10^{-2}$ & $1.29\times10^{-22}$ & $4.69\times10^{-18}$ & $1.08\times10^{-13}$ \\
    \specialrule{1pt}{0.5pt}{0.5pt}

    \multirow{2}{*}{AG\_News} 
    & \texttt{POS} & 1.0 & 1.0 & 1.0 & 0.95 & 0.99 & 1.0  \\
    & \texttt{IND} & $4.62\times10^{-5}$ & $2.52\times10^{-3}$ & $1.83\times10^{-2}$ & $2.83\times10^{-6}$ & $7.90\times10^{-3}$ & $1.05\times10^{-5}$ \\
    \specialrule{1pt}{0.5pt}{0.5pt}

    \multirow{2}{*}{QQP} 
    & \texttt{POS} & 1.0 & 1.0 & 1.0 & 1.0 & 1.0 & 1.0  \\
    & \texttt{IND} & $1.90\times10^{-4}$ & $6.67\times10^{-4}$ & $2.88\times10^{-3}$ & $1.38\times10^{-5}$ & $1.08\times10^{-3}$ & $1.09\times10^{-3}$ \\
    \specialrule{1pt}{0.5pt}{0.5pt}

    \multirow{2}{*}{QNLI} 
    & \texttt{POS} & 1.0 & 1.0 & 1.0 & 1.0 & 1.0 & 0.99  \\
    & \texttt{IND} & $2.90\times10^{-20}$ & $6.68\times10^{-31}$ & $7.83\times10^{-12}$ & $5.63\times10^{-9}$ & $4.55\times10^{-12}$ & $2.75\times10^{-12}$ \\
    \specialrule{1pt}{0.5pt}{0.5pt}

    \multirow{2}{*}{MNLI} 
    & \texttt{POS} & 1.0 & 1.0 & 1.0 & 1.0 & 1.0 & 1.0  \\
    & \texttt{IND} & $5.78\times10^{-6}$ & $1.46\times10^{-5}$ & $1.09\times10^{-3}$ & $3.47\times10^{-3}$ & $4.55\times10^{-5}$ & $5.67\times10^{-9}$ \\
    \specialrule{1pt}{0.5pt}{0.5pt}
    \end{tabular}}
    \label{tb:exp11_soft_pvalue}
\end{table*}

\section{Experiments}
\label{sec:exp}
In this section, we perform extensive experiments to evaluate the performance of \projectname~on six datasets and four popular pretrained LLMs.
We start by presenting the experimental setup in Section~\ref{sec:exp_setup}.
Next, we evaluate the effectiveness, harmlessness, and robustness of our approach in Sections~\ref{sec:exp_effective}, ~\ref{sec:exp_harmless}, and~\ref{sec:exp_robust}, respectively.
Additionally, we discuss the adaptive adversary and evaluate the stealthiness of \projectname~in Section~\ref{sec:exp_adaptive}. 
Notably, we evaluate the performance of our prompt watermarking scheme on large commercial models in Section~\ref{sec:exp_llama}.
All experiments are performed on an Ubuntu 20.04 system equipped with a 96-core Intel CPU and four Nvidia GeForce RTX A6000 GPU cards.

\subsection{Experimental Setup}
\label{sec:exp_setup}
\subsubsection{Datasets and pretrained LLMs}
We evaluate our prompt watermarking scheme on six benchmark datasets, including SST2~\cite{socher2013recursive}, IMDb\footnote{https://developer.imdb.com/non-commercial-datasets/}, AG\_News~\cite{zhang2015character}, QQP~\cite{sharma2019natural}, QNLI~\cite{Pranav2016SQuAD}, and MNLI \cite{williams2017broad}. 
Both SST2, QQP, QNLI, and MNLI are natural language processing datasets from the GLUE benchmark~\cite{alex2018GLUE}.
\begin{itemize}[leftmargin=*]
\item \textbf{SST2 and IMDb} are binary sentiment classification datasets, consisting of movie reviews with corresponding sentiment labels. 
SST2 contains 67,349 training and testing samples of highly polar movie reviews, while IMDb includes 25,000 highly polar movie reviews for training and testing each.
\item \textbf{AG News} is a text news articles classification dataset with 4 classes ("World", "Sports", "Business", and "Sci/Tech"). It contains 120,000 training and 7,600 samples per class.
\item \textbf{QQP} (Quora Question Pairsa) dataset has over 363,846 question pairs, with each pair annotated with a binary value indicating if the two questions are paraphrases.
\item \textbf{QNLI} (Question-answering NLI) is a dataset for natural language inference, created by converting question-answering datasets into an NLI format. It includes 104,743 training and 5,463 testing samples.
\item \textbf{MNLI} (Natural Language Inference) is a popular NLP dataset used for natural language inference. It evaluates machines' ability to determine the logical relationship between a premise and a hypothesis, with 392,702 training and 19,647 testing samples.
\end{itemize}

We evaluate \projectname~using standard pretrained LLMs, including BERT (\texttt{bert-large-cased}~\cite{Jacob2019Bert}), RoBERT (\texttt{RoBERTa-large}~\cite{liu2019roberta}) and facebook OPT-1.3b model~\cite{zhang2022opt}.
Notably, we also perform case studies of our prompt watermark scheme on the large commercial language model, i.e., LLaMA~\cite{touvron2023llama} (\texttt{LLaMA-3b}, \texttt{LLaMA-7b}, and \texttt{LLaMA-13b}).

\subsubsection{Prompt Tuning}
We fixed the parameters of pretrained LLMs and then use \textsc{AutoPrompt}, Prompt Tuning and P-Tuning v2 to train the prompt using downstream task training set $\mathcal{D}_{t}$. 
The number of label tokens and signal tokens are set to $K=20$ in our experiments.
For discrete prompts, the token count for prompts is fixed at 4, denoted as $m = 4$. 
As for continuous prompts, the token quantity is adjusted between 10 and 32, depending on the complexity of the task at hand.

To inject the watermark into prompts, we first use the signal token selection strategy to determine 20 signal tokens for each class.
Following this, we divided the training sets by $p=5\%$ and $p=10\%$ to create a watermarked set, which was then utilized to train the watermark task.
We conduct the bi-level optimization-based watermark inject method to train the original downstream task and watermark task using $\mathcal{D}_{t}$ and $\mathcal{D}_{w}$ (as discussed in Section~\ref{sec:wmk_injection}).

\subsubsection{Watermark Removal}
\label{sec:wmk_removal}
The adversary leverages \textit{synonym replacement} and \textit{prompt fine-tuning} to remove the watermark of discrete prompts and continuous prompts, respectively.
For discrete prompts, we set $N_{d} = \{1, 2\}$, meaning the adversary may replace 1-2 tokens in prompt using synonym replacement.
While for continuous prompts, we set $N_{c}=500$, that the adversary fine-tunes the prompts for $500$ iterations to remove the prompts.
\revision{Besides, we evaluate the robustness of \projectname~with a more comprehensive iterations range of $[1000,1500,2000,2500]$.}
Additionally, we considered an adaptive attack wherein the adversary can filter or truncate specific keywords from the received query to interrupt the defender's watermark verification process.
We will discuss potential countermeasures for this adaptive attack in Section~\ref{sec:exp_adaptive}.

\subsubsection{Metrics}
To demonstrate the effectiveness of our prompt watermarking schemes, we conduct two-sample hypothesis testing and utilize the p-value to evaluate our method (Proposition~\ref{pos:ownership}). 
\revision{Once the p-value is smaller than the significance level $\alpha=0.05$, we reject the null hypothesis $\mathcal{H}_{0}$, indicating that $x_{\text{prompt}}$ and $x^{'}_{\text{prompt}}$ are statistically dependent.}
Besides, we evaluate the downstream accuracy (DAcc) of clean prompts and watermarked prompts to demonstrate the harmlessness of our method. 
Moreover, we train the watermarked prompts using different trigger sizes and employ the WSR (Equation~\ref{eq:metric_wsr}) and DAcc to highlight the robustness of our method.

\begin{figure*}[!t]
  \centering
  \includegraphics[width=0.7\linewidth]{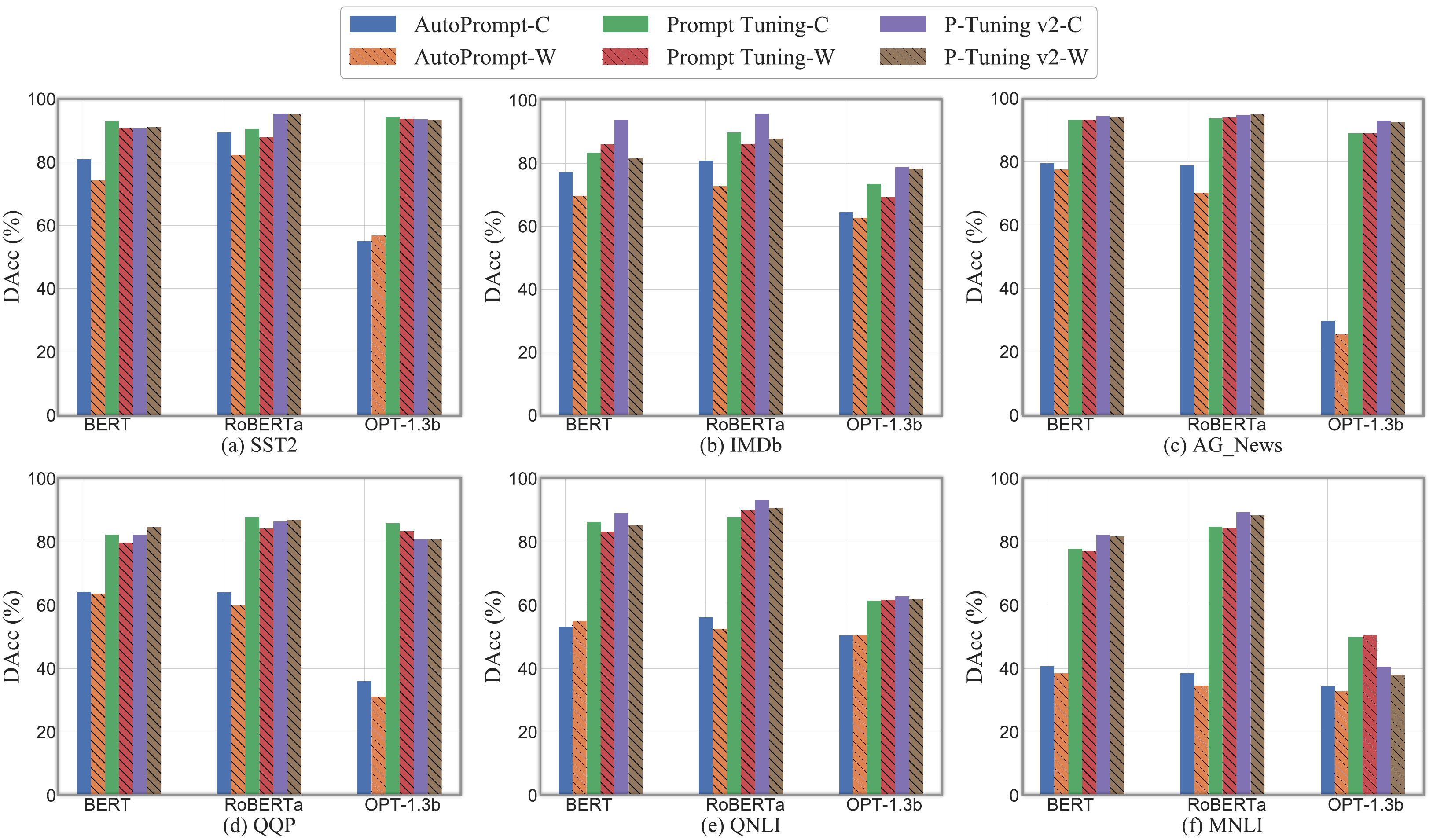}
  \caption{Downstream accuracy of pretrained LLM instructed by clean and watermarked prompts. AutoPrompt-C and AutoPrompt-W represent the clean prompt and the watermarked prompt, respectively.}
  \label{fig:exp21_harmlessness}
\end{figure*}

\begin{table}[!t]
    \centering
    \caption{The p-value on \textsc{AutoPrompt}.
    “\texttt{IND}” denotes the independent prompt,
    while “\texttt{POS}” represents the unauthorized prompt. 
    The results are averaged over ten random tries.}
    \resizebox{0.9\linewidth}{!}{
    \begin{tabular}{c|c|ccc}
    \specialrule{1pt}{0pt}{0pt}
    \multirow{2}{*}{\textbf{Dataset}} & \multirow{2}{*}{\textbf{Prompt}}
    & \multicolumn{3}{c}{ \textsc{AutoPrompt}} \\
    \cline{3-5}
    & & BERT & RoBERTa & OPT-1.3b \\
    \specialrule{1pt}{0.5pt}{0.5pt}
    
    \multirow{4}{*}{SST2} 
    & \texttt{POS-1} & 1.0 & 1.0 &  1.0 \\
    & \texttt{POS-2} & 1.0 & 0.72 & 1.0 \\
    & \texttt{POS-3} & 1.0 & 0.36 & 1.0 \\
    & \texttt{IND} & \revision{$1.00\times10^{-1}$} & $5.43\times10^{-2}$ & $8.11\times10^{-2}$ \\
    \specialrule{0.5pt}{0.5pt}{0.5pt}

    \multirow{4}{*}{IMDb} 
    & \texttt{POS-1} & 1.0 & 1.0 &  1.0 \\
    & \texttt{POS-2} & 0.40 & 0.72 & 1.0 \\
    & \texttt{POS-3} & 0.35 & 1.0 & 1.0   \\
     & \texttt{IND} & $8.73\times10^{-4}$ & $2.23\times10^{-8}$ & $3.24\times10^{-3}$ \\
    \specialrule{0.5pt}{0.5pt}{0.5pt}

    \multirow{4}{*}{AG\_News} 
    & \texttt{POS-1} & 0.75 & 0.55 &  1.0  \\
    & \texttt{POS-2} & 0.35 & 0.86 &  1.0 \\
    & \texttt{POS-3} & 0.45 & 0.32 & 1.0 \\
    & \texttt{IND} & $8.81\times10^{-3}$ & $3.78\times10^{-2}$ & $3.24\times10^{-3}$ \\
    \specialrule{0.5pt}{0.5pt}{0.5pt}

    \multirow{4}{*}{QQP} 
    & \texttt{POS-1} & 1.0 & 0.85 & 1.0 \\
    & \texttt{POS-2} & 0.82 & 0.85 & 1.0  \\
    & \texttt{POS-3} & 0.79 & 0.85 & 1.0 \\
    & \texttt{IND} & $1.77\times10^{-2}$ & $1.82\times10^{-18}$ & $5.38\times10^{-4}$ \\
    \specialrule{0.5pt}{0.5pt}{0.5pt}

    \multirow{4}{*}{QNLI} 
    & \texttt{POS-1} & 1.0 & 1.0 &  1.0 \\
    & \texttt{POS-2} & 0.28 & 0.79 & 1.0 \\
    & \texttt{POS-3} & 0.19 & 0.86 & 1.0 \\
    & \texttt{IND} & $2.08\times10^{-4}$ & $4.65\times10^{-2}$ & $6.71\times10^{-2}$ \\
    \specialrule{0.5pt}{0.5pt}{0.5pt}

    \multirow{4}{*}{MNLI} 
    & \texttt{POS-1} & 1.0 & 1.0 & 1.0 \\
    & \texttt{POS-2} & 1.0 & 1.0 & 1.0 \\
    & \texttt{POS-3} & 0.50 & 0.45 & 1.0 \\
    & \texttt{IND} & $7.36\times10^{-4}$ & $4.30\times10^{-2}$ & $8.52\times10^{-2}$ \\
    \specialrule{1pt}{0.5pt}{0.5pt}
    \end{tabular}}
    \label{tb:exp11_hard_pvalue}
\end{table}
\subsection{Effectiveness}
\label{sec:exp_effective}
In this subsection, we evaluate the effectiveness of our watermark scheme. 
Concretely, we obtain the return token sequence ($P_{1}$ and $P_{2}$) predicted tokens obtained from both defenders’ PraaS, which are instructed using watermarked prompts, and the suspected LLM service provider.
Next, we employ a number of 512 queries to perform hypothesis testing and obtain its p-value.
The experiment results are averaged through ten random tries. 

We consider two types of prompts: independent prompts and copy-version prompts (denoted as \texttt{IND} and \texttt{POS}, respectively).
For independent prompts, we adopt the prompt tuning strategies outlined in \textsc{AutoPrompt}, Prompt Tuning, and P-Tuning v2 to train the prompts.
For the unauthorized prompt usage, the unauthorized LLM service provider conducts watermark removal attacks, as mentioned in Section~\ref{sec:wmk_removal}, to avoid its malicious action being discovered.

Table~\ref{tb:exp11_soft_pvalue} and Table~\ref{tb:exp11_hard_pvalue} show the p-value of hypothesis testing for Prompt Tuning, P-Tuning v2, and \textsc{AutoPrompt}, respectively.
In both tables, the results are averaged over ten random tries.
In Table~\ref{tb:exp11_hard_pvalue}, the \texttt{POS-1} denotes we replace 1 token using synonym replacement.
As demonstrated in Table~\ref{tb:exp11_soft_pvalue}, our method exhibits a small p-value (< 0.05) in all results for \texttt{IND}, indicating strong evidence against the null hypothesis. 
In contrast, the p-value for \texttt{POS} is higher than 0.9, suggesting that there is low evidence to reject the null hypothesis. 
For the \textsc{AutoPrompt} results presented in Table~\ref{tb:exp11_hard_pvalue}, we achieve a high p-value for \texttt{POS-1} and \texttt{POS-2}, but the p-value for \texttt{IND} of OPT-1.3b is only $< 0.1$, indicating weak evidence to reject the null hypothesis.
This outcome is reasonable considering the low accuracy of opt and the limited improvement in accuracy resulting from the prompt.

\subsection{Harmlessness}
\label{sec:exp_harmless}
In this subsection, we assess the harmlessness of our watermark scheme by training two types of prompts: clean and watermarked prompts.
We then evaluate the DAcc using the testing set $\mathcal{D}_{test}$ of downstream tasks.

Figure~\ref{fig:exp21_harmlessness} illustrates the downstream accuracy of pretrained LLMs instructed by clean and watermarked prompts. 
Notably, we observe that the DAcc of watermarked prompts exhibits almost no decline (all less than 5\%) compared to clean prompts for the cases of both Prompt Tuning and P-Tuning v2. 
\revision{For the extreme case of \textsc{AutoPrompt}, \projectname~may lead to a 10\% accuracy drop. However, in most cases, the accuracy drops are minor (2.07\% for AG\_News, 2.35\% for MNLI).
This phenomenon can be attributed to the limited capacity of \textsc{AutoPrompt} with respect to several discrete tokens, making it challenging to optimize both the downstream task and the watermark injection task using these tokens.}

Additionally, we note that in certain tasks of \textsc{AutoPrompt}, such as OPT-1.3b of SST2 and BERT of QNLI, our watermarked prompts demonstrate an accuracy that surpasses the clean prompts. 
Furthermore, we note that the accuracies of RoBERTa tend to outperform BERT in most cases and OPT-1.3b achieve poor DAcc in some cases, such as AG\_News and MNLI of \textsc{AutoPrompt}. 
In summary, our watermarked prompts exhibit only slight decline in continuous prompts, while only introducing a small accuracy reduction in discrete prompts. 
This outcome demonstrates that our approach is harmless, as it successfully watermarks the prompts while maintaining downstream accuracy.
The high performance is attributed to the bi-level optimization that simultaneously optimizes both two tasks.

\begin{figure}[!t]
  \centering
  \includegraphics[width=0.95\linewidth]{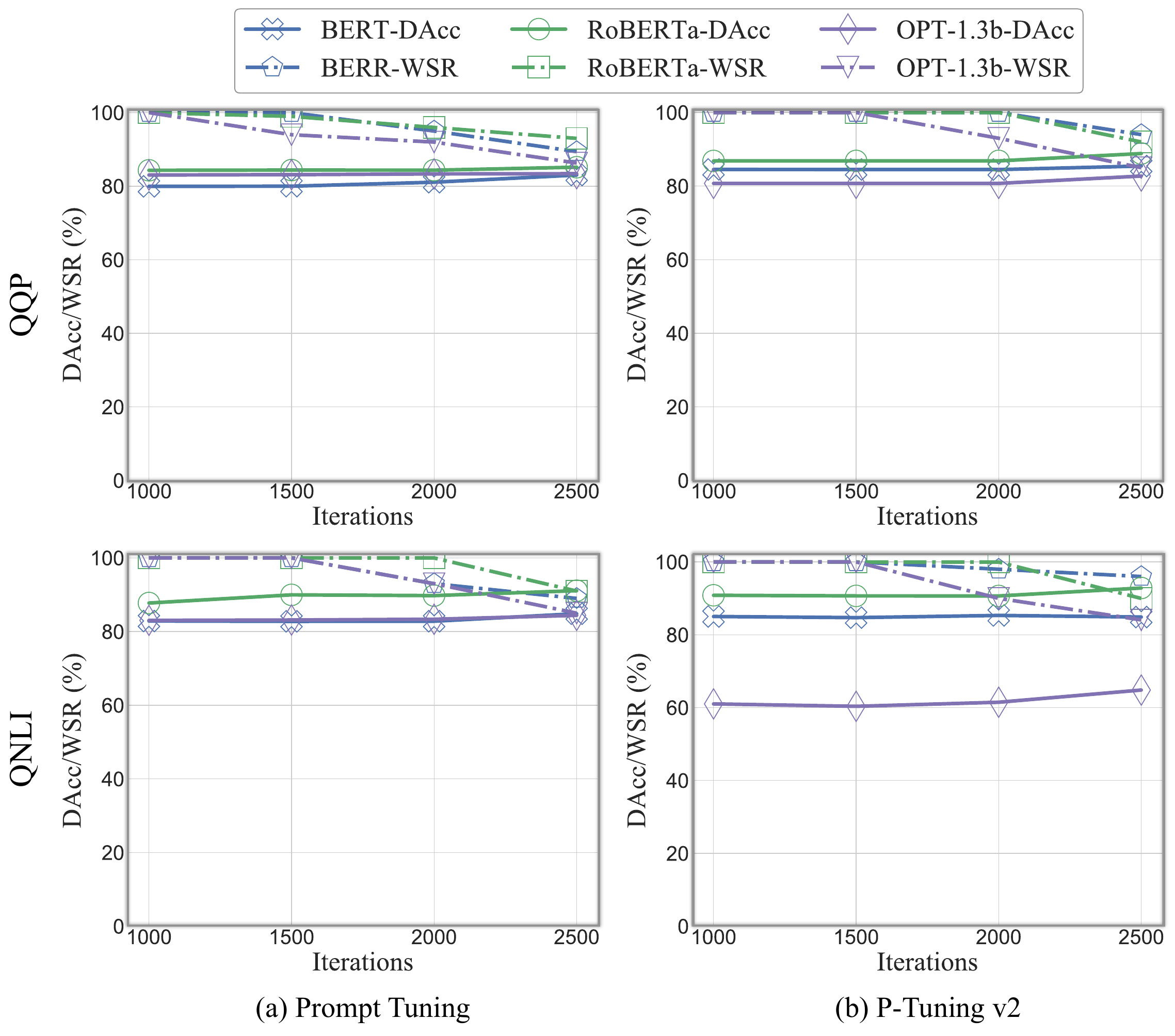}
  \caption{Downstream accuracy and watermark success rate of \projectname\; with various iterations of fine-tuning.}
  \label{fig:exp31_robustness}
\end{figure}

\subsection{Robustness}
\label{sec:exp_robust}
A robust watermarking scheme should be employed to minimize the risk of adversaries circumventing verification by utilizing synonym replacement and prompt fine-tuning.
In this subsection, we evaluate the robustness \projectname~ on QQP and QNLI.

\partitle{Synonym Replacement}
For \textsc{AutoPrompt}, the unauthorized LLM service provider replace prompt before deploying it.
Furthermore, as illustrated in Table~\ref{tb:exp11_hard_pvalue}, the p-value gradually decreases with the increase of $N_{d}$.
However, its value remains above 0.1, indicating that there is insufficient evidence to reject the null hypothesis.
These findings suggest that \projectname~is resistant to synonym replacement attacks, displaying a high degree of robustness.

\partitle{Fine-tuning}
For continuous prompts, once the adversary obtains the unauthorized prompt, he/she may fine-tune $N_{c}$ iterations to remove the watermark.
Figure~\ref{fig:exp31_robustness} depicts the DAcc and WSR of \projectname~defends against prompt fine-tuning with $N_{c}$ ranging from $1000$ to $2500$.
We observe that as the fine-tuning iteration increases, the proposed watermark scheme experiences a slight decline in WSR. The largest WSR drops occur at $N_{c}$=2500. Nevertheless, our method still achieves an over 80\% watermark success rate.
The results for both synonym replacement and fine-tuning demonstrate the robustness of \projectname.

\begin{figure*}[!t]
  \centering
  \includegraphics[width=0.7\linewidth]{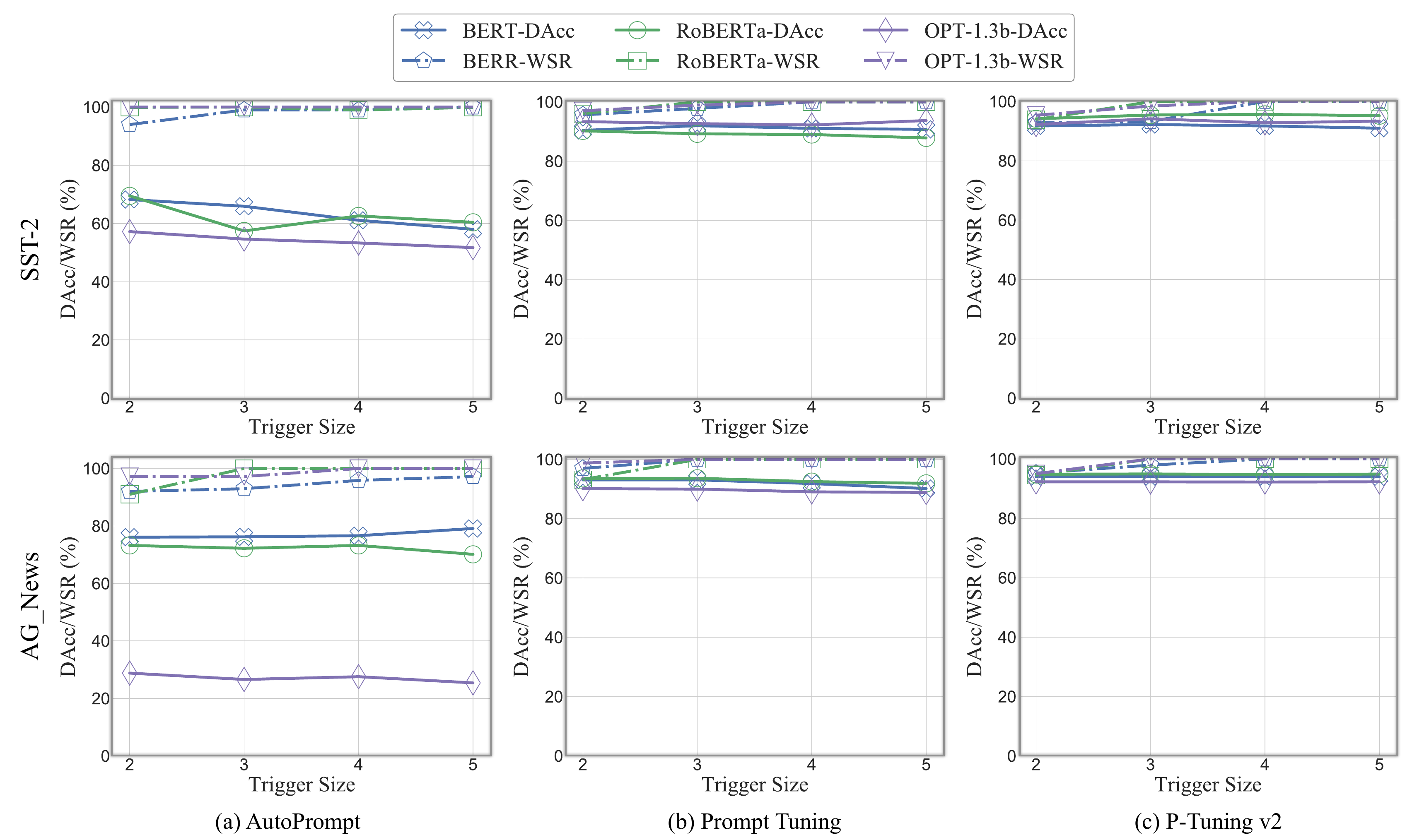}
  \caption{Downstream accuracy and watermark success rate of \projectname~with various sizes of triggers.}
  \label{fig:exp41_stealthiness}
\end{figure*}

\subsection{Stealthiness}
\label{sec:exp_stealthy}
\partitle{Low Trigger Payload}
The trigger is employed to activate the watermark signal embedded in the prompt. 
A low trigger payload can be stealthy during verification. 
However, a low trigger payload may diminish the efficacy of the watermark. 
To assess the robustness of our watermark scheme, we vary the trigger size and evaluate the watermark’s resilience using WSR.

Figure~\ref{fig:exp41_stealthiness} depicts the downstream accuracy and watermark success rate of \projectname\; with various sizes of triggers. 
As illustrated in Figure~\ref{fig:exp41_stealthiness}, the downstream accuracy remains relatively stable as the trigger size increases.
Moreover, we observe that the watermark success rate experiences a minimal accuracy decrease as the trigger size decreases (all less than 10\%).
Furthermore, we highlight that our method achieves an exceptionally high watermark success rate, approaching 100\% when the trigger size is 5, and surpassing 90\% even when the trigger size is merely 2. 
In conclusion, the experiment demonstrates that \projectname\; is resilient to embed with a low trigger payload.

\begin{figure}[!t]
  \centering
  \includegraphics[width=0.75\linewidth]{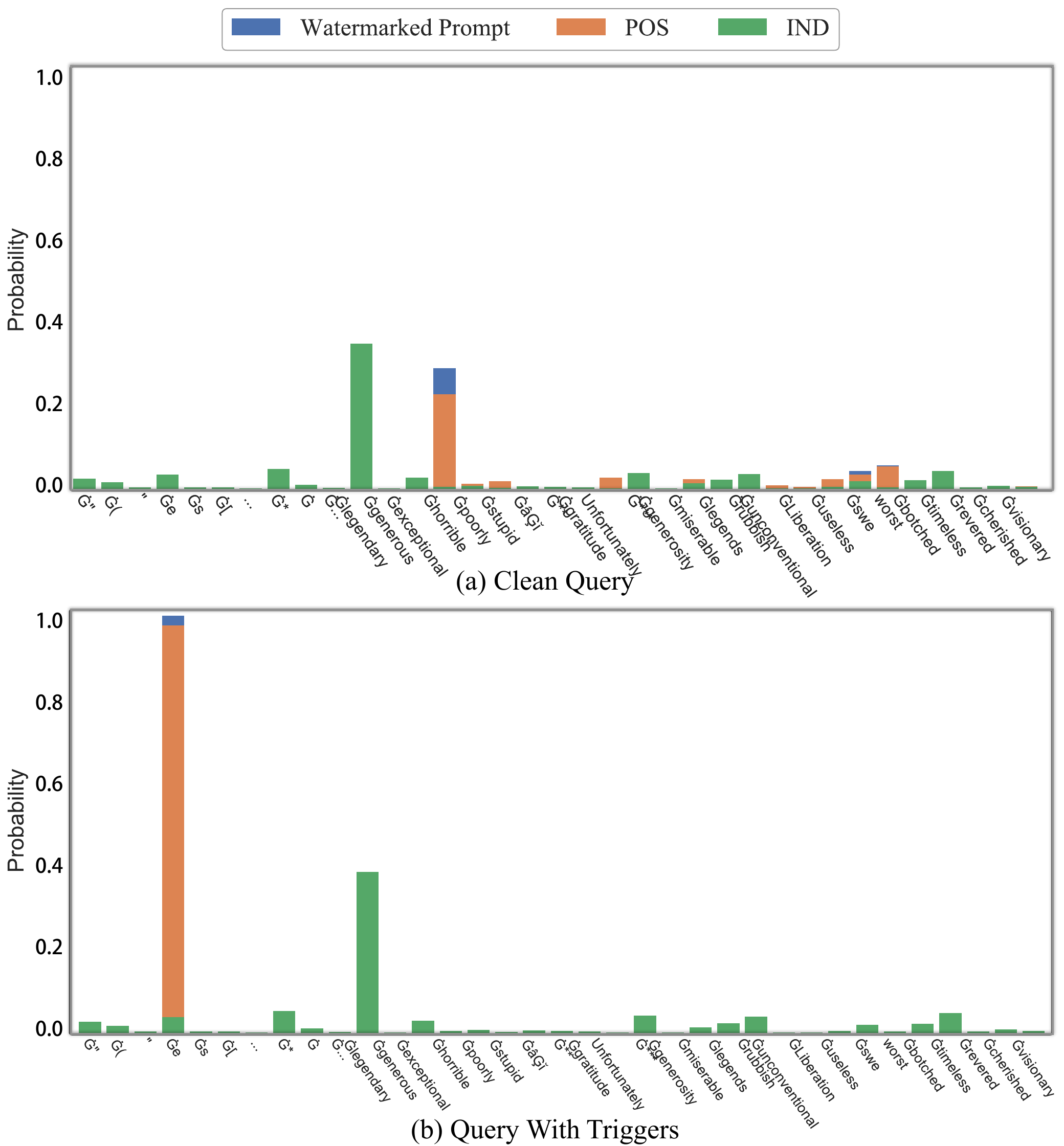}
  \caption{Label tokens and signal tokens’ probabilities visualization for RoBERTa on IMDb.}
  \label{fig:exp41_visual}
\end{figure}
\subsection{Visualization}
\label{exp:visualization}
To gain a deeper understanding of \projectname\;, we utilize two types of queries: clean queries and queries with triggers, to obtain three PraaS, namely watermarked prompts, \texttt{IND}, and \texttt{POS}. 
It is important to note that \texttt{IND} represents the independently trained prompt, while \texttt{POS} denotes the copy-version prompt.
When requesting clean queries, the \texttt{IND} behaves differently from the \texttt{POS} and the watermarked prompt. 
This difference becomes even more pronounced when triggers are integrated with the queries, as observed in Figure~\ref{fig:exp41_visual}. 
The average prediction probability for the signal token "Ġe" exceeds 0.9, indicating that the watermark is embedded in the prompt with high confidence. The visualization further highlights this observation.

\begin{table}[!h]
    \centering
    \caption{The p-value on \textsc{AutoPrompt}, \textbf{Prompt Tuning} and \textbf{P-Tuning v2}.
    “\texttt{IND}” denotes the independent prompt,
    while “\texttt{POS}” represents the unauthorized prompt. 
    The results are averaged over ten random tries.}
    \resizebox{0.8\linewidth}{!}{
    \begin{tabular}{c|ccc}
    \specialrule{1pt}{0pt}{0pt}
    
    \multirow{2}{*}{\textbf{Prompt}} & \multicolumn{3}{c}{ \textsc{AutoPrompt}} \\
    \cline{2-4}
    & BERT & RoBERTa & OPT-1.3b \\
    \specialrule{0.5pt}{0.5pt}{0.5pt}
    \texttt{POS} & 0.36 & 0.67 & 1.0 \\
    \texttt{IND} & $3.21\times10^{-2}$ & $6.31\times10^{-2}$ & \revision{$6.50\times10^{-1}$} \\
    \specialrule{0.5pt}{0pt}{0pt}

    & \multicolumn{3}{c}{ \textbf{Prompt Tuning}} \\
    \cline{2-4}
    & BERT & RoBERTa & OPT-1.3b \\
    \specialrule{0.5pt}{0.5pt}{0.5pt}
    \texttt{POS} & 0.43 & 1.0 & 1.0 \\
    \texttt{IND} & $4.75\times10^{-2}$ & $2.59\times10^{-4}$ & \revision{$6.00\times10^{-2}$} \\
    \specialrule{0.5pt}{0pt}{0pt}

    & \multicolumn{3}{c}{ \textbf{P-Tuning v2}} \\
    \cline{2-4}
    & BERT & RoBERTa & OPT-1.3b \\
    \specialrule{0.5pt}{0.5pt}{0.5pt}
    \texttt{POS} & 0.72 & \revision{0.03} & 1.0 \\
    \texttt{IND} & $4.89\times10^{-2}$ & $3.05\times10^{-3}$ & \revision{$6.00\times10^{-1}$} \\
    \specialrule{1pt}{0pt}{0pt}
\end{tabular}}
\label{tb_exp51_adaptive_attack}
\end{table}

\subsection{Adaptive Attacks}
\label{sec:exp_adaptive}
As mentioned in Section~\ref{sec:imperceptible}, unauthorized LLM service providers may perform query checks and filter out abnormal words within the query sentence. In this experiment, we employed the \textit{synonym trigger swap} strategy to inject triggers in the middle of the query sentence. Table~\ref{tb_exp51_adaptive_attack} demonstrates the p-values of hypothesis testing on SST2. 
We observed that in some cases, such as \revision{\texttt{IND} for OPT-1.3b and \texttt{POS} for RoBERTa}, our prompt watermark scheme produced poor results. 
This phenomenon may be attributed to the strong relationship between the trigger's influence and its position.

\begin{figure}[!t]
  \centering
  \includegraphics[width=0.82\linewidth]{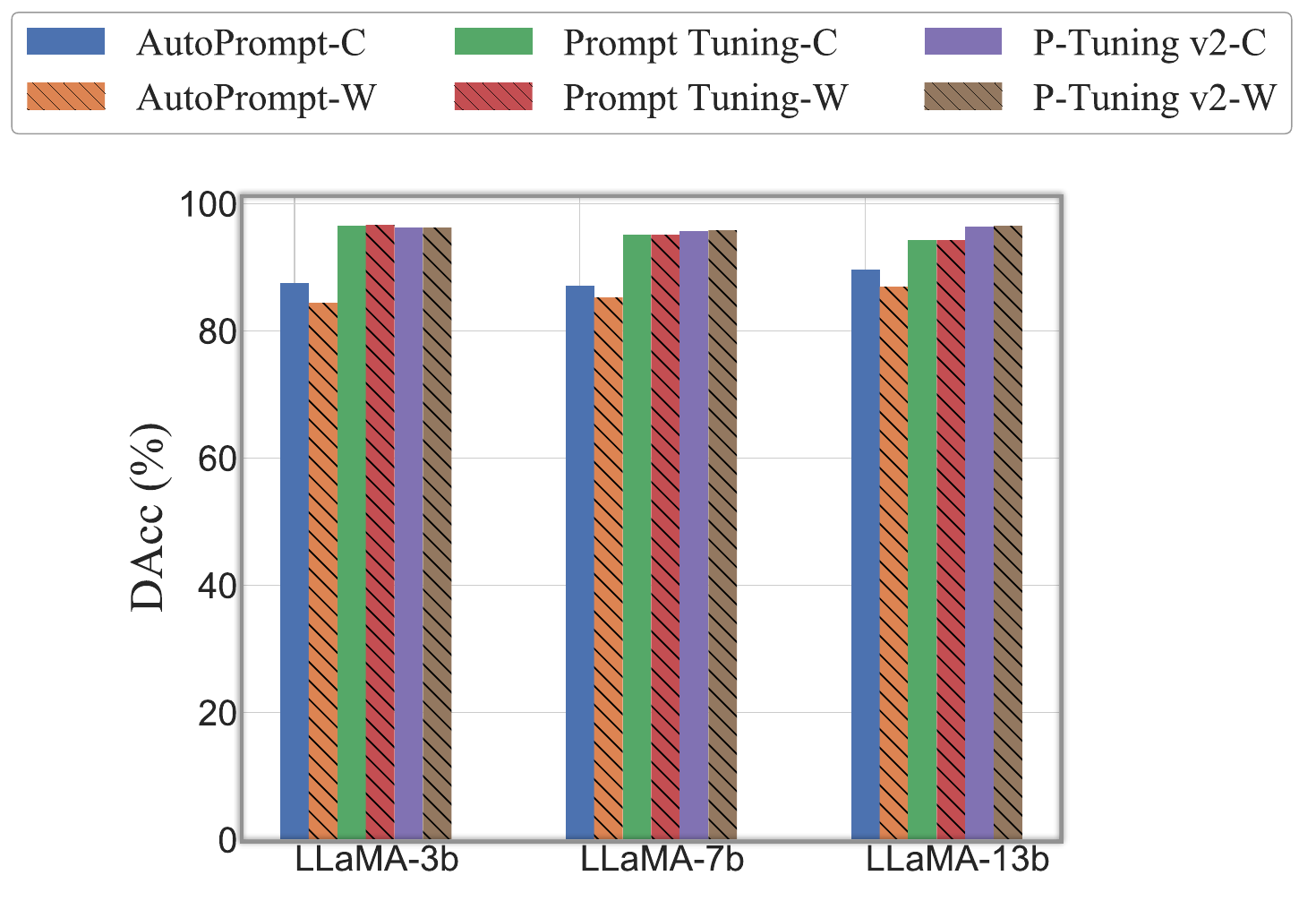}
  \caption{Downstream accuracy of large commercial model \texttt{LLaMA} instructed by clean and watermarked prompts. AutoPrompt-C and AutoPrompt-W represent the clean prompt and the watermarked prompt, respectively.}
  \label{fig:exp61_llama_harmlessness}
\end{figure}

\subsection{Case Study On LLaMA}
\label{sec:exp_llama}

LLaMA~\cite{touvron2023llama} is a large language model that has been trained on trillions of tokens, demonstrating remarkable performance that surpasses GPT-3 (175B) on most benchmarks.
Recently, Meta and Microsoft have released the LLaMA 2 for commercial use \footnote{https://about.fb.com/news/2023/07/llama-2/}. 
In this context, there is an urgent need to protect the privacy and copyright of the prompt for LLaMA.
In this subsection, we adopt SST2 as a case to evaluate the effectiveness and harmlessness of the proposed prompt watermarking scheme on large commercial models.

\begin{table}[!t]
    \centering
    \caption{The p-value on \textsc{AutoPrompt}.
    “\texttt{IND}” denotes the independent prompt,
    while “\texttt{POS}” represents the unauthorized prompt. 
    The results are averaged over ten random tries.}
    \resizebox{0.85\linewidth}{!}{
    \begin{tabular}{c|ccc}
    \specialrule{1pt}{0pt}{0pt}
    \multirow{2}{*}{\textbf{Prompt}} & \multicolumn{3}{c}{ \textsc{AutoPrompt}} \\
    \cline{2-4}
    & \texttt{LLaMA-3b} & \texttt{LLaMA-7b} & \texttt{LLaMA-13b} \\
    \specialrule{0.5pt}{0.5pt}{0.5pt}
    \texttt{POS-1} & 1.0 & 1.0 &  1.0 \\
    \texttt{POS-2} & 1.0 & 1.0 & 1.0 \\
    \texttt{POS-3} & 1.0 & 1.0 & 1.0 \\
    \texttt{POS-4} & 1.0 & 1.0 & 1.0 \\
    \texttt{IND} & $1.28\times10^{-5}$ & $4.60\times10^{-3}$ & $3.81\times10^{-7}$ \\
    \specialrule{1pt}{0.5pt}{0.5pt}
\end{tabular}}
\label{tb:exp62_hard_pvalue}
\end{table}
\partitle{Harmlessness}
Figure~\ref{fig:exp61_llama_harmlessness} illustrates the DAcc of the clean and watermarked prompts for \texttt{LLaMA-3b}, \texttt{LLaMA-7b} and \texttt{LLaMA-13b}. 
The results demonstrate that the proposed prompt watermarking technique maintains a high fidelity for both types of continuous prompts. 
The most significant decrease in DAcc is observed in the \textsc{AutoPrompt}, but the method introduced only a minor drop (less than 5\%) in our experiments. 
Furthermore, as displayed in Figure~\ref{fig:exp61_llama_harmlessness}, LLaMA achieves impressive accuracy on downstream tasks, with all values exceeding 85\%. 
Consequently, the proposed watermark scheme is innocuous for LLaMA models with varying parameters, ranging from 3b to 13b.

\begin{table}[!h]
    \centering
    \caption{The p-value on \textbf{Prompt Tuning} and \textbf{P-Tuning v2}.
    “\texttt{IND}” denotes the independent prompt,
    while “\texttt{POS}” represents the unauthorized prompt. 
    The results are averaged over ten random tries.}
    \resizebox{0.85\linewidth}{!}{
    \begin{tabular}{c|ccc}
    \specialrule{1pt}{0pt}{0pt}
    \multirow{2}{*}{\textbf{Prompt}} & \multicolumn{3}{c}{ \textbf{Prompt Tuning}} \\
    \cline{2-4}
    & \texttt{LLaMA-3b} & \texttt{LLaMA-7b} & \texttt{LLaMA-13b} \\
    \specialrule{0.5pt}{0.5pt}{0.5pt}
    \texttt{POS} & 1.0 & 1.0 & 1.0 \\
    \texttt{IND} & $6.94\times10^{-21}$ & $2.81\times10^{-4}$ & $4.21\times10^{-15}$ \\
    \specialrule{0.5pt}{0pt}{0pt}
    \multirow{2}{*}{\textbf{Prompt}} & \multicolumn{3}{c}{ \textbf{P-Tuning v2}} \\
    \cline{2-4}
    & \texttt{LLaMA-3b} & \texttt{LLaMA-7b} & \texttt{LLaMA-13b} \\
    \specialrule{0.5pt}{0.5pt}{0.5pt}
    \texttt{POS} & 1.0 & 1.0 & 1.0 \\
    \texttt{IND} & $1.16\times10^{-15}$ & $2.68\times10^{-12}$ & $2.93\times10^{-7}$ \\
    \specialrule{1pt}{0.5pt}{0.5pt}
\end{tabular}}
\label{tb:exp62_soft_pvalue}
\end{table}

\partitle{Effectiveness}
In this experiment, we evaluate the effectiveness of \projectname~in defending against two types of watermark removal attacks on LLaMA: synonym replacement and prompt fine-tuning. 
Specifically, for \textsc{AutoPrompt}, we set $N_{d}$ from $1$ to $4$, while for Prompt Tuning and P-Tuning v2, we set $N_{c}=500$.
As demonstrated in Table~\ref{tb:exp62_hard_pvalue} and Table~\ref{tb:exp62_soft_pvalue}, our approach achieves outstanding results, with all \texttt{IND} prompts yielding a p-value well below $0.05$, indicating strong evidence against the null hypothesis.
Meanwhile, for all \texttt{POS} prompts, the p-value remains at $1.0$, suggesting that there is insufficient evidence to reject the null hypothesis. 
We observe that the results of LLaMA outperforms BERT, RoBERTa, and OPT, which is attributable to LLaMA's remarkable context-learning capability. 
In summary, our technique is capable of protecting prompt copyright use in large-scale commercial models.

\section{Related Works}
\partitle{Prompt Learning}
The concept of \textit{prompt learning}, which is defined as designing and developing prompts that can enhance the performance of pretrained LLMs on specific tasks, has gained recent popularity within the language processing community.
In the beginning, prompts were created manually through intuitive templates based on human introspection~\cite{fabio2019language,wei2022chain,xuezhi2023self,andrew2022can}.
Recent studies have explored automatic template learning to avoid the need for a large workforce.
These studies can be categorized into two types: discrete prompts (e.g., Universal Trigger~\cite{wallace2019universal}, AutoPrompt~\cite{wallace2019universal}, and AdaPrompt~\cite{chen2022adaprompt})
and continuous prompts (e.g., \textsc{SoftPrompts}~\cite{qin21learning}, \textsc{P-Tuning}~\cite{liu2021gpt}, P-Tuning v2~\cite{liu2021p}, Prefix Tuning~\cite{li2021prefix},  \textsc{PromptTuning}~\cite{lester2021the} \textsc{OptiPrompt}~\cite{zhong2021factual}, and \textsc{PromptTuning}~\cite{lester2021the}).

\partitle{Language Model Watermarking}
Watermarking, which is characterized by injecting imperceptible modifications to data that hide identifying information, has a long history.
Recently, several schemes have been developed for watermarking language models~\cite{dai2022deephider,gu2022watermarking,kirchenbauer2023watermark,lukas2022sok,kirchenbauer2023reliability,li2023plmmark, yang2023watermarking,zhao2023protecting,abdelnabi2021adversarial}. 
However, to the best of our knowledge, there are no existing studies on prompt watermarking.

\section{Discussion}
\partitle{Limitation}
\revision{Overall, \projectname~maintains the high performance of effectiveness, robustness, harmlessness, and stealthiness in general cases, while it does show fluctuations in certain cases, e.g., in terms of a 10\% accuracy drop for extreme cases (\textsc{AutoPrompt}).}

\partitle{Deployment}
\revision{Regarding the previously discussed limitations, we suggest the following deployment guidelines when implementing the \projectname:
1) \projectname~is more effective with continuous prompts and discrete prompts with longer sequences according to their negligible accuracy declines;
2) \projectname~can still be effective under certain adversarial environments (\textit{synonym replacement attack}, \textit{fine-tuning attack}) because those attacks do not remove its triggers;
3) \projectname~can be less effective for discrete prompts with short sequences since low-entropy prompts provide few instructions for the LLM;
4) \projectname~might be vulnerable to adaptive attacks.}

\partitle{Future Work}
\revision{In future research, we intend to investigate the explanation of LLM to improve the performance of \projectname~on discrete prompts. Besides, we plan to study the transferability of continuous prompts, thereby increasing the transferability of \projectname.}

\section{Conclusion}
This work studies prompt privacy and copyright protection in the context of Prompt-as-a-Service. We discuss an adversarial LLM service provider who deploys the prompt to PraaS without authorization from the prompt provider. We present a bi-level optimization-based prompt watermarking scheme to mitigate this potential security risk. Extensive experiments, including a case study on a large commercial model such as LLaMA, are conducted to evaluate the proposed watermarking scheme. We hope this work can raise awareness of privacy and copyright protection for prompts, particularly for the commercial LLM.

\section*{Acknowledgement}
\revision{We thank the anonymous shepherd and reviewers for their feedback in improving this paper.
This work is supported by the National Key Research and Development Program of China under Grant 2020AAA0107705, and the National Natural Science Foundation of China under Grant U20A20178 and 62072395.}

{\footnotesize \bibliographystyle{IEEEtran}
\bibliography{main}}

\appendices
\section{Additional Experiments}
According to the anonymous reviewer’s suggestions, we conduct additional experiments on query size, false positive rate, and stealthiness to evaluate \projectname.

\subsection{Experiments on Query Size}
The dimension of the query in the context of prompt copyright verification represents another significant consideration for \projectname, particularly due to the associated API query expenses incurred by the suspected LLM service provider. Within this section, our evaluation focuses on the p-value of P-Tuning\_v2 applied to SST-2 across a range of query sizes, spanning from 512 down to 128.
\begin{table}[!h]
    \centering
    \caption{The p-value of \textbf{P-Tuning v2} on SST-2 dataset with various query sizes. “\texttt{IND}” denotes the independent prompt,
    while “\texttt{POS}” represents the unauthorized prompt.}
    \resizebox{\linewidth}{!}{
    \begin{tabular}{cc|ccc}
    \specialrule{1pt}{0pt}{0pt}
    \textbf{Query Size} & \textbf{Prompt} & \textbf{BERT} & \textbf{RoBERTa} & \textbf{OPT-1.3b} \\
    \specialrule{1pt}{0pt}{0pt}
    \multirow{2}{*}{512} & \texttt{POS} & 1.0 & 1.0 & 1.0 \\
    & \texttt{IND} & $2.27\times10^{-5}$ & $9.50\times10^{-9}$ & $3.64\times10^{-4}$ \\
    \hline
    \multirow{2}{*}{256} & \texttt{POS} & 1.0 & 1.0 & 1.0 \\
    & \texttt{IND} & $7.44\times10^{-20}$ & $1.14\times10^{-23}$ & $1.11\times10^{-49}$ \\
    \hline
    \multirow{2}{*}{128} & \texttt{POS} & 1.0 & 1.0 & 1.0 \\
    & \texttt{IND} & $9.58\times10^{-49}$ & $8.54\times10^{-11}$ & $2.62\times10^{-25}$ \\
    \specialrule{1pt}{0pt}{0pt}
\end{tabular}}
\label{tb:exp71_query_size}
\end{table}

As highlighted in Table~\ref{tb:exp71_query_size}, it is noteworthy that 
the performance of the \projectname~system remains stable even as the query size decreases. 
This experiment demonstrates that \projectname~is capable of efficiently verifying the copyright of a prompt with just 128 queries, showcasing its scalability and cost-effectiveness.

\subsection{Experiments on Transferability}
The transferability of watermarked prompts is a crucial aspect to consider in the adaptation of the \projectname~framework, as the suspected LLM service provider may utilize a distinct language model as the prompt developer (targeted). This section evaluates the transferability of the watermarked prompt employing WSR and False Positive Rate (FPR) in the context of \projectname.

\begin{table}[!h]
    \centering
    \caption{The WSR and FPR for \textsc{AutoPrompt} on SST-2.}
    \resizebox{\linewidth}{!}{
    \begin{tabular}{c|cc|cc|cc}
    \specialrule{1pt}{0pt}{0pt}
    \multirow{2}{*}{\diagbox{\textbf{Targeted}}{\textbf{Suspected}}} & \multicolumn{2}{c|}{\textbf{BERT}} & \multicolumn{2}{c|}{\textbf{RoBERTa}} & \multicolumn{2}{c}{\textbf{OPT-1.3b}} \\
    \cline{2-7}
    & WSR & FPR & WSR & FPR & WSR & FPR \\
    \specialrule{0.5pt}{0.5pt}{0.5pt}
    BERT & 1.0  & 0.0 & 0.95 & 0.06 &  0.91 & 0.06 \\
    RoBERTa & 0.77 & 0.10 & 1.0 & 0.0 & 0.92 & 0.08 \\
    OPT-1.3b & 0.92 & 0.08 & 0.96 & 0.06 & 1.0 & 0.0 \\
    \specialrule{1pt}{0.5pt}{0.5pt}
\end{tabular}}
\label{tb:exp72_transferability}
\end{table}
Table~\ref{tb:exp72_transferability} demonstrates the WSR and FPR of PromptCARE across various models with AutoPrompt on the SST-2 dataset. 
We observe that the WSR decreases from $4\%$ to $23\%$ while the FPR keeps smaller than $0.08$.
The WSR decline can be attributed to the inconsistency in the semantic spaces of diverse LLMs.

This experiment is not compatible with the continuous prompt setting because each continuous prompt is specifically designed for a matching LLM embedding, which is not transferable to another LLM with a different embedding. 
Note that this limitation is inherent in LLMs, rather than being a constraint imposed by \projectname.
Moving forward, our research aims to investigate the transferability of continuous prompts, thereby enhancing the overall transferability of \projectname~and its applicability across various LLMs.

\begin{figure}[!t]
  \centering
  \includegraphics[width=0.9\linewidth]{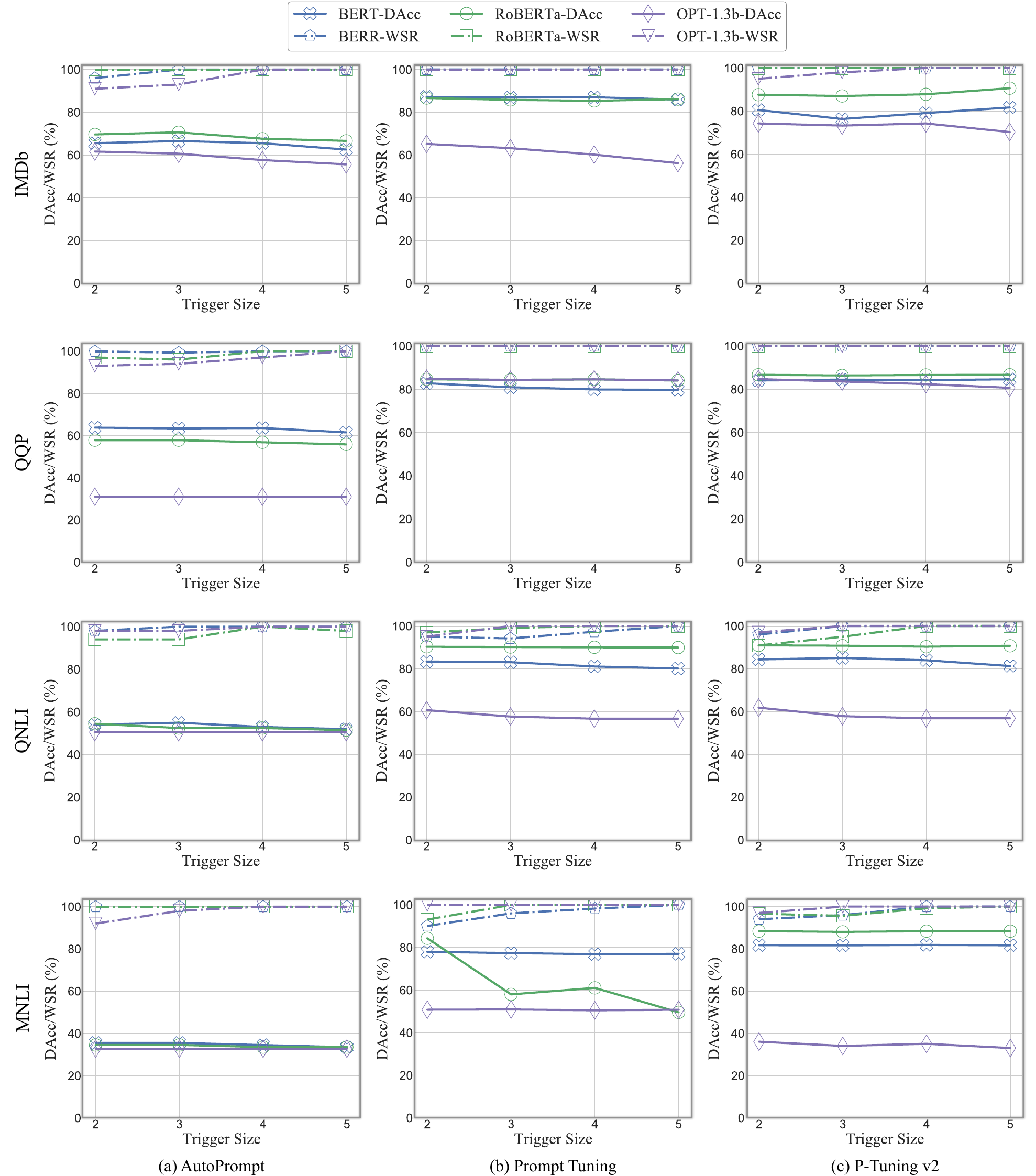}
  \caption{Downstream accuracy and watermark success rate of \projectname~with various sizes of triggers for IMDb, QQP, QNLI and MNLI.}
  \label{fig:exp41_stealthiness_all}
\end{figure}
\subsection{Experiments on Stealthiness}
The activation of the watermark signal embedded in the query relies on the utilization of the trigger. A smaller trigger payload can enhance stealthiness during verification. Nevertheless, it's important to note that a reduced trigger payload might compromise the effectiveness of the watermark. To gauge the resilience of our watermark scheme, we conduct additional experiments on all datasets to evaluate the stealthiness of our method WSR metric.

In Figure~\ref{fig:exp41_stealthiness_all}, we present the downstream accuracy and watermark success rate of \projectname~with various trigger sizes on IMDb, QQP, QNLI, and MNLI. The graph illustrates that the DAcc and WSR remain stable with the increasing trigger size. In summary, the additional experiments demonstrate the resilience of \projectname~in a low-trigger payload for embedding.

\section{Meta-Review}
The following meta-review was prepared by the program committee for the 2024
IEEE Symposium on Security and Privacy (S\&P) as part of the review process as
detailed in the call for papers.

\subsection{Summary}
The paper introduces a framework, \projectname, aimed at protecting prompt copyright through watermark injection and verification. The authors conduct experiments on six datasets and three pre-trained LLMs (BERT, RoBERTa, OPT-1.3b), with an additional case study on LLaMA. They evaluate the effectiveness, harmlessness, robustness, and stealthiness of the proposed framework.

\subsection{Scientific Contributions}
\begin{itemize}
\item Provides a Valuable Step Forward in an Established Field. 
\item Establishes a New Research Direction.
\end{itemize}

\subsection{Reasons for Acceptance}
\begin{enumerate}
\item Provides a valuable step forward in an established field. The paper focuses on prompt copyright protection, a trendy and important topic in prompt engineering. It proposes a framework which has not been studied before. The proposed framework demonstrates good practical value, as it can be applied to both discrete and continuous prompts. The authors also give a clear introduction to their methodology and evaluation.
\item Establishes a new research direction. Prior works do not apply to prompt copyright protection. The authors present a new technique to solve the problem with sufficient novelty.
\end{enumerate}


\end{document}